\documentclass[draft]{agujournal2019}
\usepackage{url} %
\usepackage{lineno}
\usepackage[inline]{trackchanges} %
\usepackage[final]{pdfpages} %
\addeditor{Ting}
\usepackage{soul}
\usepackage{amsmath}
\usepackage{amssymb}
\usepackage{tikz}
\usetikzlibrary{positioning, arrows.meta, calc, backgrounds}
\usepackage{pdflscape}
\usepackage{booktabs}

\draftfalse

\journalname{Reviews of Geophysics}

\begin{document}

\title{Anthropogenic Heat in Urban Climate Systems: Forcing, Sensitivity, and Feedback}

\authors{
Dan Li\affil{1,2},
Alvin Christopher Galang Varquez\affil{3},
Ting Sun\affil{4},
Yuya Takane\affil{5},
Jiachuan Yang\affil{6},
Mingze Ding\affil{1},
David Sailor\affil{7}
}

 \affiliation{1}{Department of Earth and Environment, Boston University, Boston, USA}
 \affiliation{2}{Department of Mechanical Engineering, Boston University, Boston, USA}
 \affiliation{3}{Department of Transdisciplinary Science and Engineering, Institute of Science Tokyo, Tokyo, Japan}
 \affiliation{4}{Department of Risk and Disaster Reduction, University College London, London, UK}
 \affiliation{5}{Center for Climate Change Adaptation, National Institute for Environmental Studies, Tsukuba, Japan}
 \affiliation{6}{Department of Civil and Environmental Engineering, Hong Kong University of Science and Technology, Hong Kong, China}
 \affiliation{7}{School of Geographical Sciences and Urban Planning, Arizona State University, Phoenix, USA}

\correspondingauthor{Dan Li}{lidan@bu.edu}

\begin{keypoints}
\item A unified forcing--response--feedback framework is developed for understanding the climatic impacts of anthropogenic heat flux.
\item The literature on anthropogenic heat flux and its climatic impacts is synthesized using this unified framework.
\item Major uncertainties and gaps in anthropogenic heat flux research are identified.
\end{keypoints}

\begin{abstract}
Anthropogenic heat flux, the heat released to the environment from human activities such as building energy use, transportation, industrial processes, and human metabolism, is a defining feature of the urban climate system. It is an important contributor to the urban heat island (UHI) effect and influences a wide range of urban meteorological processes. Its significance extends beyond urban climatology because of its close connections to energy consumption, greenhouse gas emissions, and climate adaptation. Despite decades of research, anthropogenic heat flux remains one of the least well-constrained components of the urban energy balance. Moreover, its climatic significance has often been assessed from an applied perspective, with less emphasis on developing transferable physical understanding of how anthropogenic heat flux acts as a forcing, how urban temperatures respond, and how feedbacks modify that response. This review develops a forcing--response--feedback framework for synthesizing current understanding of the role of anthropogenic heat flux in the urban climate system and identifies priorities for future research.
\end{abstract}

\section*{Plain Language Summary}

Human activities release large amounts of heat into the environment through buildings, transportation, industry, and metabolism. This anthropogenic heat contributes to local warming, influences weather and climate, and is closely linked to energy demand and climate adaptation. Despite decades of research, a unified framework for understanding its climatic effects has been lacking. This review develops a forcing–response–feedback framework showing that the climatic impact of anthropogenic heat flux depends not only on its magnitude, but also on the temperature sensitivity to this forcing and on feedbacks that modulate that sensitivity. We distinguish different definitions of sensitivity and feedback, review advances and uncertainties in estimating anthropogenic heat flux, synthesize observational and modeling evidence for both temperature sensitivity and feedback processes, and identify priorities for future research.

\section{Introduction}

Long before the emergence of contemporary climate science, scientists had already begun to consider whether the heat released from human energy consumption during urbanization and industrialization could influence weather and climate. One of the earliest discussions of this idea appeared in Luke Howard’s \textit{The Climate of London}  \cite{howard1833climate}. Howard was the first to document what is now known as the urban heat island (UHI) effect, the tendency for cities to be warmer than their surrounding rural areas, and to postulate that the UHI effect was associated with the heat released from human metabolism, domestic heating, industrial activity, and combustion sources throughout the city. Howard even remarked that “the real matter of surprise, when we contemplate so many sources of heat in a city is, that the effect on the thermometer is not more considerable.” The heat sources identified by Howard are now commonly referred to collectively as anthropogenic heat flux ($Q_F$). Throughout this review, the terms anthropogenic heat flux, anthropogenic heat emissions, anthropogenic heat release, anthropogenic heating, and, where context is clear, simply anthropogenic or waste heat are used interchangeably.

The earliest quantification of anthropogenic heat flux was perhaps provided by \citeA{eaton_1878}, who presented what he termed a “rudimentary calculation” based primarily on coal consumption for London. In modern units, Eaton’s estimate corresponds to an anthropogenic heat flux of approximately 12 W m$^{-2}$ averaged over the metropolitan district of London (about 306 km$^2$). Eaton further estimated that this heat release would produce a warming rate of approximately $1.2$ K per hour. Additional contributions from other anthropogenic heat sources, including human metabolism, increase the warming rate to  $1.4$ K per hour. Such a warming rate, if sustained over a day, would produce an urban–rural temperature difference of more than 30 K, far exceeding those observed. This quantitatively illustrates Howard’s earlier surprise that the urban temperature increase relative to its rural counterpart was “not more considerable,” and implies that heat-removal processes must offset much of the warming effect of anthropogenic heat. %

Later studies began to develop more systematic estimates of anthropogenic heat flux. \citeA{schmidt_1917}, for example, refined Eaton’s early estimate by combining energy consumption data with population-based estimates of metabolic heat to calculate spatially averaged anthropogenic heat fluxes for Vienna, Austria, and Berlin, Germany. Building on this line of work, \citeA{garnett_1965} estimated that anthropogenic heat flux in Sheffield, UK was approximately 19 W m$^{-2}$ in 1952 when averaged over its built-up urban and industrial land (about 48 km$^2$), nearly 1/5 of the incoming solar radiation and 1/3 of the net radiation; \citeA{bach_1970} estimated that anthropogenic heat flux in Cincinnati, USA, was about 26 W m$^{-2}$ in summer when averaged over an area of the city size (about 200 km$^2$). A major focus of these early studies was to compare anthropogenic heat flux against the natural components of the surface energy balance, particularly radiative fluxes, and to establish anthropogenic heat flux as an important component of the urban surface energy balance \cite{oke_1974_wmo}.

During this period, observations of urban temperatures and other meteorological variables became increasingly available, revealing systematic differences between urban and rural climates \cite <e.g.,>{landsberg1956,landsberg1970}. In particular, the UHI effect was firmly established as one of the most prominent and extensively studied urban climate phenomena \cite{lowry1967}. These studies made clear that the UHI effect could not be explained by a single mechanism, but instead emerged from the combined influences of multiple physical processes \cite{oke_1974_wmo}. This recognition motivated the development and application of numerical models of land-surface and boundary-layer processes to cities in order to disentangle the relative contributions of these different mechanisms through controlled experiments \cite <e.g.,>{myrup_1969,atwater1971,atwater1972,gutman1975,yu1975,torrance1976}.

Progressively, anthropogenic heat flux was recognized not only as an urban-scale perturbation, but as a potential influence on regional and global climate. The landmark Study of Man’s Impact on Climate (SMIC) report reflected this broader perspective \cite{smic1971}. In addition to noting the importance of anthropogenic heat flux at local and city scales, the SMIC report emphasized the strong scale dependence of anthropogenic heat fluxes, recognizing that “any system for averaging of the artificial heat input (or anthropogenic heat flux) into the Earth’s atmospheric system over large inhomogeneous areas will give only very approximate results.”  The report also highlighted emerging efforts to investigate the climatic effects of anthropogenic heat flux using general circulation models \cite{washington_1971, washington_1972}, noting that “the effects of urban and regional heat sources on continental and global climate are uncertain.”

Since then, substantial advances have been made in both the estimation of anthropogenic heat flux and the assessment of its climatic impacts. Early inventory-based approaches relying primarily on fuel consumption and energy use statistics have gradually evolved into methods based on increasingly sophisticated urban energy models, energy balance approaches, remote sensing approaches, and, more recently, machine learning methods \cite{sailor_methods,feng_2025_review}. At the same time, numerical studies of the climatic effects of anthropogenic heat flux have progressed from one-dimensional surface energy balance and boundary-layer models to two-dimensional and three-dimensional hydrostatic or non-hydrostatic atmospheric models \cite{lu_2024_review}. These studies have examined the impacts of anthropogenic heat flux across a wide range of spatial scales, from within individual urban street canyons to mesoscale urban environments and even the global climate system. Moreover, early recognition of the interactions between urban climate, building energy use, and anthropogenic heat flux \cite{nicol1976anthropogenic} has been followed by the development of coupled urban climate--building energy models (BEMs) \cite <e.g.,>{Kikegawa2003AE,Salamanca2010TAC}, enabling these feedbacks to be explicitly represented and quantified.

These advances have established several broad characteristics of anthropogenic heat flux and its climatic significance \cite{oke2017}. Physically, anthropogenic heat flux arises from the combustion of fossil and biomass fuels, nuclear reactions, the dissipation of electricity and mechanical energy to heat, and human metabolism. It is commonly attributed to three source sectors, namely, buildings, transportation, and human metabolism \cite{oke2017}, although industry is also often treated as a separate sector \cite{sailor_methods}; among these, building energy use is often a dominant contributor in dense urban areas \cite{sailor_methods}. Anthropogenic heat is also released outside cities, for example from power plants, industrial facilities, and transportation networks, but its magnitude is typically highest in urban areas. %
The magnitude of $Q_F$ at the neighborhood scale ranges from a few W m$^{-2}$ in suburban or low-density urban areas to several hundred W m$^{-2}$ in dense urban cores, with even higher localized values in highly urbanized districts; it also exhibits strong diurnal and seasonal variability, reflecting daily traffic cycles and seasonal heating and cooling demand \cite{oke2017}. Anthropogenic heat flux affects a wide range of urban meteorological variables, including temperature  \cite <see reviews in e.g.,>{Wang2023ERL,Xie2024SCS}, humidity \cite <e.g.,>{wang2018effects}, winds  \cite <e.g.,>{xie2016modeling,zhang2016numerical,ko2026modeling}, heat stresses \cite<e.g.,>{Takane2020ERC,huang2022street}, boundary-layer height \cite<e.g.,>{lin2008urban}, clouds \cite <e.g.,>{kanda2001numerical,chen2019seasonal}, and precipitation \cite <e.g.,>{holst2016sensitivity,nie2017impacts,fung2021comparing,kim2021impacts}. Focusing on temperatures, studies have documented the effects of anthropogenic heat flux on road surface temperatures \cite <e.g.,>{khalifa2016accounting,colas2025traffic}, bulk urban land surface temperatures \cite <e.g.,>{zhao2014strong,meng2016quantifying,meng2017mitigating,ueyama2020cooling,guo2024surface}, near-surface air temperatures within the urban canopy layer \cite <e.g.,>{biggart2021modelling,chen2024modelling,Li2024Structural}, and air temperatures extending above the urban canopy into the urban boundary layer \cite <e.g.,>{krpo2010impact}. The warming effects of anthropogenic heat flux are generally strongest during nighttime and wintertime conditions, when background radiative forcing is weaker and atmospheric mixing is reduced \cite<e.g.,>{atwater1972,Wang2023ERL}. As a result, anthropogenic heat flux has been identified as a key contributor to nocturnal and wintertime UHIs. 
There is also growing recognition of positive feedbacks in which urban warming increases cooling energy demand and anthropogenic heat flux in summer \cite<e.g.,>{Takane2019NPJ,Li2024NatClimate}, thereby further amplifying urban temperatures and the associated heat and energy risks.

Despite these advances, many of the central questions remain unresolved. Anthropogenic heat flux remains one of the least constrained terms in the urban energy balance. Large uncertainties persist not only in its magnitude, spatial distribution, and temporal variability, but also in its meteorological and climatic impacts. Although the literature contains a large number of studies, much of it is organized around particular cities, episodes, and modeling configurations. This reflects the applied contexts in which much of the literature has developed, but also complicates the synthesis of prior findings into more general physical understanding. Therefore, rather than reviewing the literature from the perspective of applied urban meteorology, this review adopts a broader climate science perspective and reinterprets prior work using the language of climate forcing, sensitivity, and feedback, with a specific focus on the sensible component of anthropogenic heat flux and its influence on urban temperatures. Urban temperatures and UHIs are also influenced by many other factors, such as surface characteristics, building properties, moisture fluxes, and atmospheric aerosols and greenhouse gases \cite{oke2017}, which are beyond the scope of this review. Likewise, although part of the anthropogenic heat flux is released as latent heat, for example through cooling towers \cite{sailor_methods,oke2017}, its effects on atmospheric moisture are not considered here.

Accordingly, the emphasis of this review is not on the completeness of the bibliography, but on synthesizing previous studies within a unified framework that brings applied urban meteorology into closer dialogue with climate science. To facilitate the synthesis, we use large-language-model tools to help extract and organize information from a corpus of over 500 publications related to anthropogenic heat flux. This exercise is documented in the Supporting Information, and the resulting corpus is intended as a resource for readers interested in the broader literature beyond the studies discussed in the main text. The main text only focuses on selected studies that best illustrate the climate science perspective taken in this review. The basic premise is that the climatic significance of anthropogenic heat flux depends not only on the forcing itself, but also on how urban temperatures respond to that forcing and how feedbacks amplify or damp that response. To this end, Section~2 develops a forcing–response–feedback framework that serves as the organizing structure for the review.

\section{Theory} \label{sect:theory}

Anthropogenic heat flux is commonly incorporated into the urban surface energy balance as an additional source term:
\begin{equation}
Q^\ast + Q_{F} = Q_H + Q_E + Q_S + \delta Q_A,
\label{eq:SEB}
\end{equation}
where $Q^\ast$ is net all-wave radiation, $Q_H$ and $Q_E$ are the turbulent sensible and latent heat fluxes, respectively, $Q_S$ is the heat storage flux, and $\delta Q_A$ represents net advective heat exchange. Although this equation is often referred to as an urban surface energy balance equation, it actually represents the energy balance of an urban control volume, including both urban surfaces and the overlying atmosphere, rather than that of an individual urban facet (e.g., roofs, walls, and ground surfaces), as emphasized by \citeA{oke2017}. In this formulation, anthropogenic heat flux is treated analogously to an additional radiative energy input into this urban control volume.

From the perspective of understanding the climatic effects of anthropogenic heat flux, the manner in which anthropogenic heat flux enters the urban system is also important. %
Early numerical studies already recognized this issue. For example, \citeA{myrup1970corrigendum} noted that simply adding anthropogenic heat flux to the natural net radiation term as in Eq. \ref{eq:SEB} may be overly simplistic because the thermal effects of anthropogenic heat flux depend on where the anthropogenic heat is released and how it interacts with atmospheric stratification and turbulent transport \cite<see also>{KondoKikegawa2003,Kikegawa2003AE}. More recently, \citeA{Li2024Structural} showed that the release location of anthropogenic heat flux in numerical models can alter its thermal effects by as much as an order of magnitude.

Here we treat anthropogenic heat flux as an energy input to a control volume of urban canopy air. Our goal is not to argue whether this is the optimal representation of anthropogenic heat flux in the urban climate system, but rather to use the heat budget of the canopy air as a conceptual basis for illustrating our forcing--response--feedback framework. %
Within the confines of this treatment, at the simplest level, anthropogenic heat flux acts as a \textit{forcing} that perturbs the canopy air temperature ($T$), with the resulting temperature change referred to as the \textit{response}. However, the temperature response to anthropogenic heat flux is not determined solely by the imposed forcing because both the processes governing energy exchange and the anthropogenic heat flux itself may depend on canopy air temperature, thereby creating \textit{feedbacks}.

A central quantity in this forcing–response–feedback framework is the \textit{sensitivity} of urban temperature to anthropogenic heat flux, which quantifies the  temperature change produced by a unit perturbation in anthropogenic heat flux and is referred to as \textit{anthropogenic heat sensitivity} throughout this review. For consistency, we define this sensitivity using canopy air temperature in this review, but it could be extended  using other temperature metrics. Following the climate change convention (e.g., \citeNP{roe2009}), consider a perturbation in anthropogenic heat flux $\Delta Q_F$ applied to an initial equilibrium atmospheric state, producing a new equilibrium state with the canopy air temperature change between these two equilibrium states of $\Delta T_{eq}$, where the subscript `eq' indicates equilibrium. Here and throughout this paper, $\Delta$ denotes a change or perturbation in a quantity. The anthropogenic heat sensitivity is then defined as
\begin{equation}
S \equiv \frac{\Delta T_{eq}}{\Delta Q_{F}},
\label{eq:S}
\end{equation}
with units of K (W m$^{-2}$)$^{-1}$.

\subsection{A canopy air energy balance model}
\label{sec:energy-balance}

For a volume of  air with mass $m$ (kg), density $\rho$ (kg m$^{-3}$), and heat capacity $c_p$ at constant pressure (J kg$^{-1}$ K$^{-1}$) within an urban canopy of horizontal area $A$ (m$^2$) and depth $h$ (m), the canopy air temperature tendency (${dT}/{dt}$, K s$^{-1}$) is governed by
\begin{equation}
m c_p \frac{dT}{dt} = A Q = A (Q_{F} + Q_{other}),
\end{equation}
where $Q$ is the net energy flux (W m$^{-2}$) that includes the anthropogenic heat flux $Q_{F}$ and other energy fluxes  collectively represented by $Q_{other}$. Dividing by area gives
\begin{equation}
C \frac{dT}{dt} = Q_{F} + Q_{other},
\label{eq:canopy_air_energy_balance}
\end{equation}
where $C = {m c_p}/{A} = {\rho c_p h}$ is an effective heat capacity per unit area (J K$^{-1}$ m$^{-2}$).
For simplicity, $C$ will be treated as a constant.

If $Q_{F}$ is a constant and $Q_{other}=0$, temperature would increase linearly with time. One of the earliest examples of this reasoning was provided by \citeA{eaton_1878}, as discussed in the Introduction. With $Q_F = 12\ \mathrm{W\,m^{-2}}$ and assuming $h = 30 $ m, Eaton estimated a warming tendency of approximately $1.2\ \mathrm{K\,h^{-1}}$ (or $2.1^\circ\mathrm{F\,h^{-1}}$, as reported in Eaton’s original paper).

In contrast to this limiting case, the presence of $Q_{other}$ may counteract the warming induced by anthropogenic heat flux, thereby preventing unbounded temperature increase. As the simplest case, $Q_{other}$ is assumed to be a net heat-loss term represented as a linear function of $T$, namely, $Q_{other}=-(\lambda T + c)$, where $\lambda$ is the slope (W m$^{-2}$ K$^{-1}$) assumed to be positive,  $c$ is the intercept (W m$^{-2}$). Under this convention, the negative sign reflects the tendency for the net heat-loss term to become more negative as canopy air temperature increases. With this linear assumption, the energy balance model reads
\begin{equation}
C\frac{dT}{dt}=Q_{F}-(\lambda T+c).
\label{eq:linear-energy-balance}
\end{equation}

\begin{figure}[htbp]
    \centering
    \includegraphics[width=\textwidth]{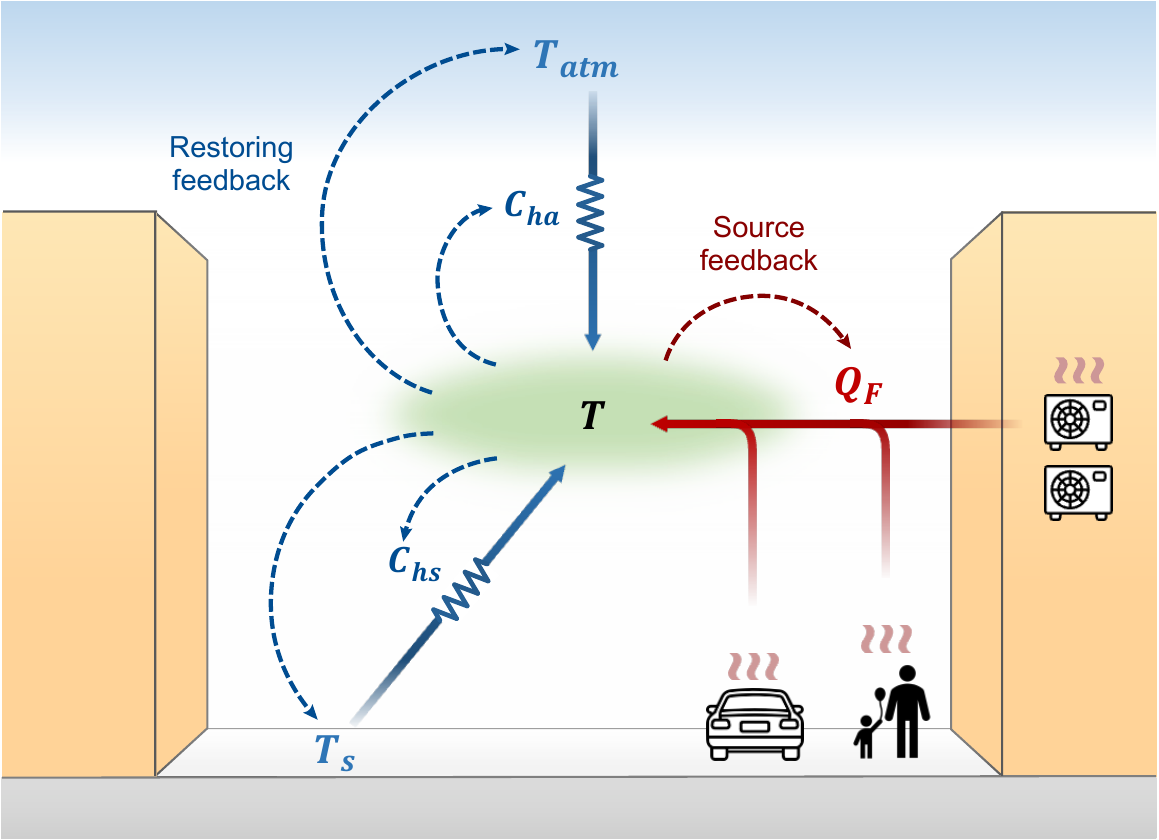}
    \caption{A schematic illustration of the forcing–response–feedback framework. Anthropogenic heat flux ($Q_F$) acts as a forcing that warms the urban canopy air temperature ($T$). The canopy air exchanges heat with urban surfaces, represented for simplicity by a single effective surface temperature $T_s$ (see Section~\ref{sec:energy-balance}), and the overlying atmosphere, represented by atmospheric temperature $T_{atm}$; $C_{hs}$ and $C_{ha}$ denote the canopy air–surface and canopy air–overlying atmosphere heat conductances, respectively. Restoring feedbacks arise when these heat conductances, or the surface and atmospheric temperatures, vary with canopy air temperature. At the same time, urban warming may increase anthropogenic heat flux through source feedback, for example by increasing building cooling demand. The diagram is used to illustrate the canopy air energy balance model used in this review, emphasizing heat transfer processes commonly represented in urban canopy parameterizations rather than including all physical processes that may influence canopy air temperature. Solid arrows indicate heat fluxes, with both $Q_F$ and $Q_{other}$ (Eq.~\ref{eq:bulk-formulation}) defined as positive when they represent energy input to the canopy air; dashed arrows indicate feedbacks.}
    \label{fig:theory}
\end{figure}

Eq. \ref{eq:linear-energy-balance} describes a forced--damped  system in which anthropogenic heat flux drives warming while heat-loss processes provide a restoring tendency. The linear model for $Q_{other}$ is not required by the forcing--response--feedback framework (Section~\ref{subsect:interpreting-framework}); it is adopted here because it is consistent with the formulations widely used in urban canopy parameterizations within numerical weather and climate models (see Figure \ref{fig:theory} for a schematic), where $Q_{other}$ is often expressed as the sum of canopy air--overlying atmosphere turbulent exchange and canopy air--surface heat exchange (involving turbulent and conductive processes), both parameterized using bulk-transfer formulations:
\begin{equation}
Q_{other}
=
\rho c_p C_{ha}(T_{atm}-T)
+
\rho c_p C_{hs}(T_s-T).
\label{eq:bulk-formulation}
\end{equation}
Here $C_{ha}$ and $C_{hs}$ are canopy air--overlying atmosphere and canopy air--surface heat conductances (m s$^{-1}$), respectively; $T_{atm}$ is the air temperature above the canopy layer, and $T_s$ is the urban surface temperature. This is consistent with the linear model for $Q_{other}$ with
\begin{equation}
\lambda=\rho c_p(C_{ha}+C_{hs}),
\label{eq:lambda conductances}
\end{equation}
and
\begin{equation}
c=-\rho c_p(C_{ha}T_{atm}+C_{hs}T_s).
\label{eq:c conductances}
\end{equation}
For simplicity, Eq.~\ref{eq:bulk-formulation} represents all urban surfaces by a single effective surface temperature $T_s$. In reality, and in many numerical models, several surfaces with different surface temperatures, such as roofs, sunlit and shaded walls, and pervious and impervious ground, exchange heat with the canopy air. When considering multiple surfaces, the second term on the right-hand side of Eq.~\ref{eq:bulk-formulation} then becomes an area-weighted sum over these surfaces \cite<e.g.,>{Wang2023ERL}, but the framework developed below still applies.

\subsection{A forcing--response--feedback framework}
\label{subsect:framework}

The energy balance model (Eq. \ref{eq:linear-energy-balance}) can be treated in two different ways depending on whether feedbacks are included or not. In the reference case, $Q_{F}$ is a prescribed forcing and the parameters in the energy balance model (e.g., $\lambda$ and $c$) do not depend on $T$. When feedbacks are included, the temperature response modifies either the forcing itself or the heat-loss mechanisms.

\subsubsection{Reference  sensitivity}
\label{sec:open-loop}

The simplest solution and interpretation of Eq. \ref{eq:linear-energy-balance} is obtained by neglecting feedbacks and treating $Q_{F}$, $C$, $\lambda$ and $c$ as constants and thus independent of $T$.
In this case, Eq. \ref{eq:linear-energy-balance} has an analytical solution
\begin{equation}
T(t)=T_{eq}+\left[T(0)-T_{eq}\right]\exp\left(-\frac{t}{\tau}\right),
\end{equation}
where
\begin{equation}
T_{eq}=\frac{Q_{F}-c}{\lambda}
\label{eq:equilibrium}
\end{equation}
is the equilibrium temperature and
\begin{equation}
\tau=\frac{C}{\lambda}
\end{equation}
is the adjustment timescale. $T(0)$ is the initial canopy air temperature at $t = 0$. In this case, the temperature does not increase indefinitely, but instead relaxes exponentially toward $T_{eq}$ with the adjustment timescale $\tau$. One can now see that $C$ represents the thermal inertia of the urban canopy air layer and sets the adjustment timescale of the system. A quick estimate of $\tau$ suggests that it is on the order of 100--1000~s for an urban canyon, corroborated by simulations in \citeA{mei2021}. This short timescale is partly why weather and climate models tend to assume equilibrium for the canopy air energy balance (i.e., completely ignoring the effect of $C$). 

In the following, we will only focus on the equilibrium  sensitivity and thus drop the subscript `eq' in the definition of anthropogenic heat sensitivity $S$. It follows from Eq. \ref{eq:equilibrium} that $\Delta Q_{F}$ produces a corresponding equilibrium temperature change of $\Delta T =\Delta Q_{F}/ \lambda $. The anthropogenic heat sensitivity is therefore
\begin{equation}
S_0 \equiv \left.\frac{\Delta T}{\Delta Q_{F}}\right|_{ref} = \frac{1}{\lambda_0}.
\label{eq:S0}
\end{equation}
Following the terminology from the global climate change literature, we refer to $S_0$ as the \textit{reference sensitivity}, where the subscript zero emphasizes that no other parameters in the system are allowed to vary with $T$. Under these conditions, the only effect of  $\Delta Q_{F}$ is to increase $T$ until the temperature-dependent heat-loss change ($-\lambda \Delta T$) balances the imposed forcing perturbation.

If $Q_{other}$ is represented using  bulk-transfer formulations (Eq. \ref{eq:bulk-formulation}), then $\lambda = \rho c_p (C_{ha} + C_{hs})$ is controlled primarily by the efficiency of canopy air--surface and canopy air--overlying atmosphere heat exchange (Eq. \ref{eq:lambda conductances}). Using representative values of $\rho \sim 1$ kg m$^{-3}$, $c_p \sim 10^3$ J kg$^{-1}$ K$^{-1}$, and $C_{ha} + C_{hs} \sim 0.1$ m s$^{-1}$, $\lambda$ is on the order of $100$ W m$^{-2}$ K$^{-1}$. This implies a reference sensitivity on the order of $0.01$ K (W m$^{-2}$)$^{-1}$, consistent with the typical magnitude of sensitivity reported in the literature \cite{Wang2023ERL}. This also provides a simple explanation for the reported diurnal and seasonal variations in anthropogenic heat sensitivity \cite{Wang2023ERL}. Because turbulent exchanges are generally stronger during daytime and summer conditions, $\lambda$ tends to be larger during these periods, thereby reducing anthropogenic heat sensitivity. Conversely, nighttime and wintertime conditions are associated with smaller $\lambda$ values and therefore enhanced anthropogenic heat sensitivity. This further explains why anthropogenic heat sensitivity is typically stronger under calmer and more weakly ventilated conditions \cite{Wang2023ERL}. 

Previous studies that formulated canopy air heat budgets could derive a similar reference sensitivity under comparable assumptions. However, these studies did not always explicitly isolate or report this quantity, and even if they had, the exact analytical expressions would differ because the canopy air heat budget was formulated differently, particularly with respect to horizontal advection and canopy air–surface heat exchange \cite{xue2020impact,yuan2020mitigating,mei2021}. These studies generally diagnose temperature responses through numerical simulations rather than analytically separating the reference sensitivity from feedbacks, which are discussed below in Section \ref{sec:closed-loop}.

\subsubsection{Climate feedback parameters}
\label{sec:closed-loop}

The reference case assumes fixed anthropogenic heat flux as forcing and fixed heat-loss parameters (i.e., $\lambda$ and $c$), while retaining the baseline restoring mechanism. In reality, feedbacks may arise through modifications to either the atmospheric response to the forcing or the forcing itself. Here, we distinguish between these two classes of feedbacks. The first class, termed \textit{restoring feedback}, modifies the anthropogenic heat sensitivity by altering the way heat is removed from the urban canopy air. Such feedbacks emerge when $\lambda$ and $c$ depend on temperature, for example through changes in turbulent transport. The second class consists of feedbacks that act directly on the anthropogenic heat flux itself, termed \textit{source feedback}. For example, higher outdoor temperatures can increase cooling energy demand and the associated anthropogenic heat release during summer.

These feedbacks can be incorporated into the energy balance framework by allowing both the anthropogenic heat forcing and the heat-loss parameters to depend on temperature. Let $(T_0, Q_{F,0})$ denote an equilibrium state of Eq.~(\ref{eq:linear-energy-balance}), so that $Q_{F,0} = \lambda_0\, T_0 + c_0$, where $Q_{F,0} = Q_{F}(T_0)$, $\lambda_0= \lambda(T_0)$, $c_0= c(T_0)$. Consider a small perturbation to the temperature-independent/background component of anthropogenic heat flux, denoted as $\Delta Q_{F,0}$, causing a small temperature perturbation $\Delta T$,  and linearize each $T$-dependent quantity around $T_0$:
\begin{equation}
Q_{F}(T) \approx Q_{F,0} +   \lambda_\text{source}\, \Delta T + \Delta Q_{F,0} , \qquad
\lambda(T) \approx \lambda_0 + \lambda'\, \Delta T, \qquad
c(T) \approx c_0 + c'\, \Delta T,
\label{eq:linearise}
\end{equation}
where $\lambda_\text{source} \equiv \left.\partial Q_{F}/\partial T\right|_{T_0}$ is the partial derivative of $Q_{F}$ with respect to $T$ at $T_0$ (units W~m$^{-2}$~K$^{-1}$), and $\lambda' \equiv \left.\partial \lambda/\partial T\right|_{T_0} $ and $c' \equiv \left.\partial c/\partial T\right|_{T_0}$ are the partial derivatives of $\lambda$ and $c$ with respect to $T$ at $T_0$, respectively. Note that the total change in anthropogenic heat flux between the two equilibrium states (i.e., $Q_{F}(T) - Q_{F,0}$) includes the imposed forcing ($\Delta Q_{F,0}$) and the source-feedback contribution ($\lambda_\text{source}\, \Delta T$). Substituting into Eq.~(\ref{eq:linear-energy-balance}), using the equilibrium condition at $T_0$, and discarding terms quadratic in $\Delta T$, the following anthropogenic heat sensitivity can be obtained
\begin{equation}
S_f \equiv \frac{\Delta T}{\Delta Q_{F,0}} = -\frac{1}{\Lambda}.
\label{eq:sensitivity-with-feedback}
\end{equation}
where
\begin{equation}
\Lambda
=
\underbrace{(-\lambda_0)}_{\text{baseline}}
\;+\;
\underbrace{\lambda_\text{restoring}}_{\substack{\text{restoring}\\\text{feedback}}}
\;+\;
\underbrace{\lambda_\text{source}.}_{\substack{\text{source}\\\text{ feedback}}}
\label{eq:feedback-parameters}
\end{equation}
Here, $\lambda_\text{restoring}=    -\lambda' T_0-c'$. For $S$, a subscript `$f$' is added to emphasize that this anthropogenic heat sensitivity parameter is defined with respect to an imposed  forcing ($\Delta Q_{F,0}$), termed \textit{forcing-based sensitivity} hereafter. The forcing--response--feedback framework derived here is linear: it assumes small perturbations about a reference state. The limitations of this assumption are discussed in Section~\ref{subsect:beyond-linear-feedback-theory}.

In climate feedback terminology, each term on the right-hand side of Eq. \ref{eq:feedback-parameters} may be interpreted as a climate feedback parameter and $\Lambda$ as the total feedback parameter with units of W m$^{-2}$ K$^{-1}$. Eqs. \ref{eq:sensitivity-with-feedback} and \ref{eq:feedback-parameters} are written in a way that the sign of each feedback parameter in Eq. \ref{eq:feedback-parameters} directly indicates the nature of the feedback: positive values correspond to positive feedbacks that amplify the baseline temperature response, whereas negative values correspond to negative feedbacks that damp it. 

Similar to the convention of referring to the Planck response as the Planck feedback in global climate feedback analysis, we refer to $(-\lambda_0)$ as the baseline feedback parameter, although it represents the intrinsic restoring tendency of the system to remove heat from the urban canopy air layer rather than a formal feedback induced by changes in the system state. It is therefore negative by design. The remaining feedback parameters quantify feedback-induced modifications to this baseline restoring tendency. A canonical example of positive feedback arises from summer air-conditioning demand, for which $\lambda_\text{source}>0$ because cooling energy consumption and the associated anthropogenic heat increase with outdoor temperature. In contrast, a canonical example of negative feedback occurs when heat-loss processes become more efficient as the canopy air temperature increases, as may occur for positive $\lambda’$ through enhanced turbulent transport under increasingly unstable stratification.

Two limiting cases are worth noting. When all feedback parameters vanish, $\Lambda=-\lambda_0$ and the reference  sensitivity in Eq.~(\ref{eq:S0}) is recovered. On the other hand, as the positive feedback increasingly offsets the negative baseline restoring, $\Lambda$ approaches zero and the sensitivity is greatly amplified. Such high-sensitivity regimes may emerge during heatwaves or weakly ventilated nighttime conditions in dense urban environments with strong air-conditioning feedbacks.

Before closing this section, it is useful to define an \textit{effective sensitivity} based on the total realized change in anthropogenic heat flux, $Q_F(T)-Q_{F,0}$, which again includes both the forcing component ($\Delta Q_{F,0}$) and the source-feedback contribution ($\lambda_\text{source}\, \Delta T$), as follows:
\begin{equation}
S_{e} \equiv \frac{\Delta T}{Q_{F}(T)-Q_{F,0}}.
\label{eq:effective-sensitivity}
\end{equation}
This definition is particularly relevant when the  forcing component cannot be cleanly separated from source-feedback contribution (e.g., when part or all of anthropogenic heat fluxes are generated interactively). Under such conditions, the diagnosed sensitivity usually corresponds to the effective anthropogenic heat sensitivity $S_e$. Recovering $S_f$ requires estimating the anthropogenic heat flux that would occur in the absence of source feedback. However, there is no generally accepted approach for constructing this counterfactual estimate at present, and some implementations may raise energy-conservation concerns \cite{Kikegawa2022AE}. Throughout this review, forcing-based sensitivities ($S_f$) and effective sensitivities ($S_e$) are therefore distinguished whenever possible.

\subsubsection{Interpreting the forcing--response--feedback framework}
\label{subsect:interpreting-framework}

\paragraph{Source feedback and gain factor.}

The present framework recovers earlier gain-factor formulations \cite{GinzburgDemchenko2017,ginzburg2019,Kikegawa2022AE} that have been used to describe source feedback associated with building energy consumption. Within the present notation and without considering restoring feedback (i.e., $\lambda' = c' =0$), the temperature response reduces to $\Delta T = \Delta T_0/(1-g_A)$, where $\Delta T_0=S_0\Delta Q_{F,0}$ is the reference or baseline temperature response, and $g_A \equiv \lambda_\text{source} S_0 \equiv \lambda_\text{source} /\lambda_0$
is the gain factor associated with source feedback, which determines the amplification of the reference temperature response arising from temperature-dependent anthropogenic heat flux. Hence, the present framework generalizes these earlier gain-factor formulations by incorporating both source and restoring feedbacks in a unified manner.

\paragraph{Restoring feedback decomposition.}

The separation of $\lambda_\text{restoring}$ into the two terms involving $\lambda'$ and $c'$ is specific to the linear model for $Q_{other}$ adopted here for illustration, and the individual terms should be interpreted with caution. Using Eqs.~\ref{eq:lambda conductances} and \ref{eq:c conductances}, $-\lambda' T_0 = -\rho c_p\,(C_{ha}'+C_{hs}')\,T_0$ and $-c' = \rho c_p\,(C_{ha}'T_{atm}+C_{hs}'T_s+C_{ha}T_{atm}'+C_{hs}T_s')$, where primes denote derivatives with respect to $T$ at $T_0$. Both terms involve derivatives of the heat conductances multiplied by absolute temperatures ($T_0$, $T_{atm}$, or $T_s$). The two terms therefore depend on the origin of the temperature scale. Even with a physically natural origin, such as absolute zero in the Kelvin scale, the two terms are typically large and of opposite sign and largely cancel, so their individual values are difficult to interpret physically. Regrouping them gives
\begin{equation}
\label{eq:restoring-regrouped}
\lambda_\text{restoring}=\rho c_p\left[C_{ha}'\,(T_{atm}-T_0)+C_{hs}'\,(T_s-T_0)+C_{ha}\,T_{atm}'+C_{hs}\,T_s'\right].
\end{equation}
In this regrouped form, temperatures enter the conductance feedbacks (first two terms) only as differences and the temperature feedbacks (last two terms) only as derivatives. Each term is therefore independent of the temperature origin. This is the decomposition used by \citeA{Wang2023ERL}. Regardless of how it is decomposed, the total restoring feedback does not depend on the origin of the temperature scale. In this review, we therefore focus on the total restoring feedback $\lambda_\text{restoring}$.

\paragraph{Further generalization.}

When $Q_{other}$ is not represented by a linear model or when it includes processes other than those considered here, the baseline feedback parameter can be interpreted as the local sensitivity of $Q_{other}$ to canopy air temperature, $\partial Q_{other}/\partial T$, evaluated at the reference state and with other variables held fixed. The restoring feedback is then given by the difference between the total temperature derivative and this baseline restoring, $\lambda_\text{restoring}={dQ_{other}}/{dT}-{\partial Q_{other}}/{\partial T}$. With the linear model adopted here, these expressions reduce to $\partial Q_{other}/\partial T = -\lambda_0$ and $\lambda_\text{restoring}=-\lambda'T_0-c'$, respectively, consistent with the derivations presented earlier. 

Although the present review has focused on the canopy air heat budget, the same forcing--response--feedback framework can be applied to other temperature variables and control volumes governed by different energy budgets. For example, \citeA{khanh2025impact} treated the urban boundary layer as a well-mixed control volume and assumed that outgoing longwave radiation is the only explicit heat-loss mechanism (i.e., $Q_{\mathrm{other}}=-\varepsilon\sigma T^4$), where $\varepsilon$ is the atmospheric emissivity, $\sigma$ is the Stefan--Boltzmann constant, and $T$ is the boundary-layer temperature. At equilibrium, linearizing the emitted longwave radiation about a reference state yields a baseline feedback parameter of ${\partial Q_{\mathrm{other}}}/{\partial T}=-4\varepsilon\sigma T_0^3$ and a corresponding baseline sensitivity of $(4\varepsilon\sigma T_0^3)^{-1}$. This example illustrates that the analytical expressions for the baseline sensitivity and feedback parameters depend on the selected control volume and the heat-loss mechanisms included in its energy budget, whereas the underlying forcing--response--feedback framework remains unchanged.

\paragraph{Relation to the global climate framework} 

The framework developed here deliberately parallels the forcing--feedback framework used to analyze global climate sensitivity \cite{roe2009}. In that framework, the global-mean surface temperature responds to a radiative forcing, the Planck response provides the baseline restoring, and feedbacks associated with water vapor, clouds, the lapse rate, surface albedo, and others modify the response \cite{sherwood2020assessment}. In the present framework, the baseline feedback parameter ($-\lambda_0$) plays the role of the Planck response, and $\lambda_\text{restoring}$ that of the physical climate feedbacks. The source feedback ($\lambda_\text{source}$) is analogous to feedbacks in which the forcing agent itself responds to temperature, such as carbon-cycle feedbacks \cite{friedlingstein2006climate}.

Despite this formal similarity, the two frameworks differ fundamentally in how the forcing and the temperature are defined. In the global framework, the forcing is a change in the radiative flux at the top of the atmosphere and the temperature is the global-mean surface temperature. In the present framework, the forcing is the anthropogenic heat released into the urban canopy air and the temperature is the local canopy air temperature. As a result of these differences, several feedbacks familiar from the climate change literature, such as those associated with water vapor, clouds, and snow and ice cover, do not appear explicitly in our framework. They are not excluded, however: they are implicitly included in $\lambda_\text{restoring}$ to the extent that they modify $T_s$, $T_{atm}$, $C_{ha}$, or $C_{hs}$. Because some of these feedbacks evolve over time scales much longer than that of the canopy air, the extent to which they are captured in a diagnosed sensitivity depends on the time scale considered, as discussed further in Section~\ref{subsect:scale-dependence}.

This formalism provides the organizing principle for the remainder of the review. Section 3 examines $Q_F$ broadly as an urban climate forcing. Although the theoretical framework developed here distinguishes imposed forcing from source feedback, Section 3 does not separate these contributions, but instead focuses on the magnitude, spatial and temporal variability, scale dependence, global representation, and uncertainty of $Q_F$ as a heat input to the urban climate system. Section \ref{sect:sensitivity} examines the anthropogenic heat sensitivity by synthesizing reported $S_f$ and $S_e$ across observations and numerical models. Section \ref{sect:feedback}      examines the restoring and source feedbacks that modulate urban temperature responses.

\section{Forcing}

Early reviews established the foundational methods for estimating $Q_F$, including energy-consumption inventories, residual surface energy balance approaches, and BEMs \cite{sailor_methods}. More recent reviews have synthesized the expanding literature on $Q_F$ estimation, including the advances and challenges associated with these foundational methods, as well as the growing use of remote sensing data and machine learning techniques \cite{lu_2024_review,feng_2025_review}. Building on these syntheses, this section focuses on what recent studies reveal about anthropogenic heat as an urban climate forcing, with particular attention to its magnitude, spatiotemporal variability, scale dependence, and uncertainty.

\subsection{From bulk estimates to structured forcing}

The variability and scale dependence of $Q_F$ have long been recognized \cite<see e.g.,>{smic1971}. The global, annual mean $Q_F$ is generally on the order of $0.01$ W m$^{-2}$. However, this aggregation masks its strong concentration in urban cores, industrial districts, transport corridors, and dense building clusters; it also obscures strong diurnal, weekly, seasonal, and weather-related variability associated with commuting, occupancy, appliance use, heating, ventilation, and air conditioning (HVAC) operation. At finer spatial and temporal scales, reported values can be several orders of magnitude larger. For example, the estimated peak $Q_F$ values of 1590 W m$^{-2}$ from a single grid cell in Shinjuku, central Tokyo during winter \cite{ichinose1999}, and up to 6000 W m$^{-2}$ over the footprint of a single data center building in Phoenix, Arizona \cite{sailor_datacenter}, are among the highest reported in the literature. Recent advances in estimating $Q_F$ can be broadly viewed as efforts to characterize it as a spatially heterogeneous, temporally variable, and sectorally structured urban climate forcing.

\subsubsection{Spatial heterogeneity}

Earlier inventories often disaggregated energy statistics using population or land-use proxies, whereas more recent studies increasingly combine geospatial analysis techniques with remote sensing products to better resolve intra-urban heterogeneity. These products include nighttime light data \cite{chen_2016,lin_2020}, active fire data for detecting heavy emissions from industrial plants \cite{china_industry_2019}, and ground- or satellite-based pollutant emissions records \cite{pollutant_reg_2014}. As discussed further in Section \ref{sec:global_datasets}, global datasets have increasingly adopted data-fusion approaches that combine satellite products with sociodemographic information \cite{datafusion_2017}. Machine-learning techniques, such as cubist and random forest algorithms, have further extended this line of research \cite{china_cubist_2020,shanghai_cubist_2024,he_2023,lin_2020}. These developments have substantially improved the spatial detail of $Q_F$ estimates, but they also introduce uncertainties associated with the selection, quality, representativeness, and transferability of the predictors used to spatially disaggregate $Q_F$.

Remote sensing has also been used to estimate $Q_F$ directly. Satellite-based residual surface energy balance approaches using multispectral sensors (e.g., Landsat) have been applied to estimate spatially varying $Q_F$ over Fuzhou \cite{fuzhou_landsat_2013}, Delhi \cite{delhi_landsat_2015}, Athens and Paris \cite{mitraka_2017}, six US cities \cite{zhou_2021}, and a variety of urban environments \cite{landsat_fluxes_2025}. In these studies, $Q_F$ is estimated as the residual of the urban surface energy balance equation (i.e., Eq. \ref{eq:SEB}), so uncertainties in the estimated radiative, sensible, latent, and ground heat fluxes propagate directly into the inferred anthropogenic heat flux. In addition, because the estimates are usually tied to clear-sky satellite overpasses, they provide only snapshots in time and cannot resolve the full diurnal cycle or variability associated with changing weather conditions \cite{chrysoulakis_2018,landsat_fluxes_2025}.

\subsubsection{Temporal variability}

The temporal variability of $Q_F$ has long been recognized, with diurnal, weekly, and seasonal cycles reflecting traffic activity, weekday--weekend differences, occupancy patterns, and heating and cooling demand \cite{oke2017}. Beyond these established patterns, two further temporal dynamics have received growing attention: (1) short-term lags between building energy consumption and anthropogenic heat release, and (2) long-term changes in both past and future anthropogenic heat fluxes.

Building energy consumption and anthropogenic heat release are not always temporally coincident because of thermal storage. For a simulated building, \citeA{Liu2022ACP} showed that the hourly ratio of actual anthropogenic heat release to energy consumption can vary from negative values to well above unity. Yet energy consumption is still widely used as a proxy for $Q_F$, so this lag is neglected in many datasets. Even when the daily total is unchanged, such shifts in the timing of heat release can alter the temperature response, because anthropogenic heat sensitivity varies over the diurnal cycle (Section~\ref{subsect:numerical}).

At longer time scales, studies have estimated long-term increases in global $Q_{F}$ through past, middle, and late twenty-first century \cite{lu_2017,pf_ahf_2019,varquez_2021}. Regional to local scenario studies increasingly examine how development pathways, building efficiency, air-conditioning adoption, and transportation electrification alter future $Q_F$ \cite{darmanto_2019,ribeiro_2021}. These projections, which are designed to be consistent with integrated climate change scenarios (e.g., Representative Concentration Pathways, Shared Socioeconomic Pathways), emphasize that future $Q_F$ estimates should not be inferred from population growth alone, but also from changes in background climate, cooling/heating demand, technological advancements influencing energy use efficiency, and other socioeconomic factors.

\subsubsection{Sector-specific forcing}

Another recent development is the increasingly explicit representation of sector-specific anthropogenic heat fluxes. BEMs coupled with urban land surface and atmospheric models have enabled dynamic estimates of anthropogenic heat fluxes from the building sector under different climate, retrofit, and urban development scenarios \cite{crawley_2001_energyplus,singh_singapore_2022}. Transportation has evolved from simple traffic-volume-based estimates toward trajectory-resolved and model-integrated representations, allowing assessment of emerging technologies such as vehicle electrification \cite{singapore_ev_2020,mussetti2022electric,ChenYang2022SCS,Hata2025ACP,beijing_trajectory_2022,colas2025traffic,Sun2026JAMES}. Recent studies have also begun to isolate high-intensity point sources, including industrial facilities and data centers \cite{sailor_datacenter}, which may contribute little to city-wide averages but can generate substantial localized heat emissions \cite{lee_1979,dc_trends_2021,singh_2022,varquez_2021}.

Sector-specific estimates have improved the partitioning of $Q_F$ among buildings, transportation, industry, and other sources. However, partitioning by sector is not the same as representing the forcing pathway. For example, $Q_F$ released at street level by vehicles, ejected from HVAC systems at varying building facades, or emitted from industrial stacks can interact with the urban climate system in fundamentally different ways. This makes source-pathway representation a major unresolved challenge in anthropogenic heat research.

\subsection{Global datasets}  \label{sec:global_datasets}

\begin{landscape}

\begin{table}[p]
\centering
\scriptsize
\caption{Comparison of global anthropogenic heat flux ($Q_F$) datasets.}
\label{tab:global_qf_comparison}

\begin{tabular}{p{2.4cm} p{1.6cm} p{1.5cm} p{1.5cm} p{3.2cm} p{2.6cm} p{1.6cm} p{2.2cm} p{2.8cm}}
\toprule

\textbf{Dataset}
&
\textbf{Spatial Resolution}
&
\textbf{Period}
&
\textbf{Temporal Resolution}
&
\textbf{Methodology}
&
\textbf{Sectoral Decomposition}
&
\multicolumn{3}{c}{\textbf{Reported $Q_F$ statistics}}

\\
\cmidrule(lr){7-9}

& & & & & &
\textbf{Averaging Basis}
&
\textbf{Mean $Q_F$ ($\mathrm{W\,m^{-2}}$)}
&
\textbf{Max $Q_F$ ($\mathrm{W\,m^{-2}}$)}

\\
\midrule

Flanner2009 \cite{flanner_2009}
&
0.5$^\circ$ (finest)
&
2005--2040 (annual), 2100
&
Hourly, seasonal
&
Disaggregation of national energy statistics via population density; diurnal weightings and latitude-sinusoid temporal profiles
&
Aggregate energy use
&
Whole Earth
&
0.028 (2005); 0.059 (2040); 0.19 (2100)
&
48 (New York, $0.5^\circ$ resolution)
\\

LUCY \cite{lucy_2011}
&
2.5 arc-min
&
Contemporary
&
Hourly
&
Sectoral disaggregation (buildings, transport, metabolism) with country-specific work schedules and degree-day climatology
&
Buildings, transport, metabolism
&
Urban
&
0.7--3.6 (typical urban daytime min and max)
&
577 (New York); 262 (Paris); 178 (Tokyo)
\\

Chen and Shi (2012) \cite{chen_2012}
&
30 arc-sec
&
2006
&
Annual
&
DMSP/OLS nighttime lights calibrated with state-level energy statistics
&
Aggregate energy use
&
Land surface
&
0.10
&
$>$100
\\

Dong et al. (2017) \cite{dong_2017}
&
30 arc-sec
&
Contemporary
&
Hourly
&
Sectoral disaggregation of energy statistics using population density, nighttime lights, empirical temperature-energy relations, and temporal templates
&
Buildings, industry, transport, metabolism
&
Land surface
&
0.13
&
493 (Hong Kong); 353 (Singapore)
\\

AH-DMSP (2017) \cite{ah_dmsp_2017}
&
1 km
&
1992--2010
&
Annual
&
Statistical modeling (nighttime lights, NDVI, population, EIA statistics) of long-term annual trends
&
Aggregate energy use
&
Not reported
&
Not reported
&

581.3 (unspecified grid)
\\

PF-AHF (2019) \cite{pf_ahf_2019}
&
30 arc-sec
&
1970--2050
&
Annual
&
Population-density disaggregation of BP energy statistics
&
Aggregate energy use
&
Land surface
&
0.05 (1970); 0.13 (2015); 0.16 (2050)
&
Not reported
\\

AH4GUC \cite{varquez_2021}
&
1 km
&
2010s, 2050s
&
Hourly
&
Extension of \citeA{dong_2017} incorporating VIIRS datasets and future projections 
&
Buildings, industry, transport, metabolism
&
Whole Earth
&
0.031 (2010s); 0.050 (2050s)
&
Not reported
\\

Wang et al. (2022) \cite{wang_2022}
&
500 m
&
2016
&
Monthly
&
Multi-source data fusion with local temperature degree-day scaling
&
Building, industry, transport, metabolism
&
Land surface
&
0.19
&
438 (London); 406.4 (New York); 356.5 (Tokyo)
\\
\bottomrule
\end{tabular}

\vspace{0.5em}

\begin{minipage}{0.97\linewidth}
\footnotesize
\textit{Notes.} Temporal coverage refers to the years represented by each dataset, whereas temporal resolution denotes the frequency at which anthropogenic heat fluxes are reported. Mean $Q_F$ values are reported using different spatial averaging operators (whole Earth, land surface, or urban areas) and are therefore not directly comparable. Max $Q_F$ values correspond to the largest values explicitly reported in each study and may represent regional means, grid-cell maxima, or threshold values.
\end{minipage}
\end{table}

\end{landscape}

\citeA{sailor_methods} concluded by calling for standardized anthropogenic heat flux products for atmospheric modeling, analogous to global land-use and land-cover forcing datasets. Over the past decade, this vision has largely been realized through several global and continental-scale datasets that now provide $Q_F$ forcing for regional and global climate models. These datasets have greatly expanded the accessibility of $Q_F$ information, particularly for data-sparse regions, and enabled systematic intercomparison across cities and climate zones.

Table~\ref{tab:global_qf_comparison} highlights eight such datasets, revealing a clear trend toward increasingly resolved representation of $Q_F$. Spatial resolution has improved from 0.5$^\circ$ to 500 m, allowing increasingly detailed depiction of urban hotspots. Temporal representation has likewise evolved from annual means, to resolving seasonal and daily variations, and to hourly estimates incorporating sector-specific and weather-dependent temporal profiles. At the same time, methodologies have progressed from simple population-based disaggregation of national energy statistics toward increasingly sophisticated approaches that incorporate nighttime lights, sector-specific information, and, more recently, multi-source data fusion. However, the increased realism does not necessarily imply reduced uncertainty. Differences in spatial resolution, temporal representation, source allocation, and downscaling strategy can produce substantially different $Q_F$ estimates. Consequently, although most datasets reproduce similar large-scale geographical patterns, they can diverge markedly in the magnitude and distribution of $Q_F$ within individual cities.

\begin{figure}[htbp]
    \centering
    \includegraphics[width=\textwidth]{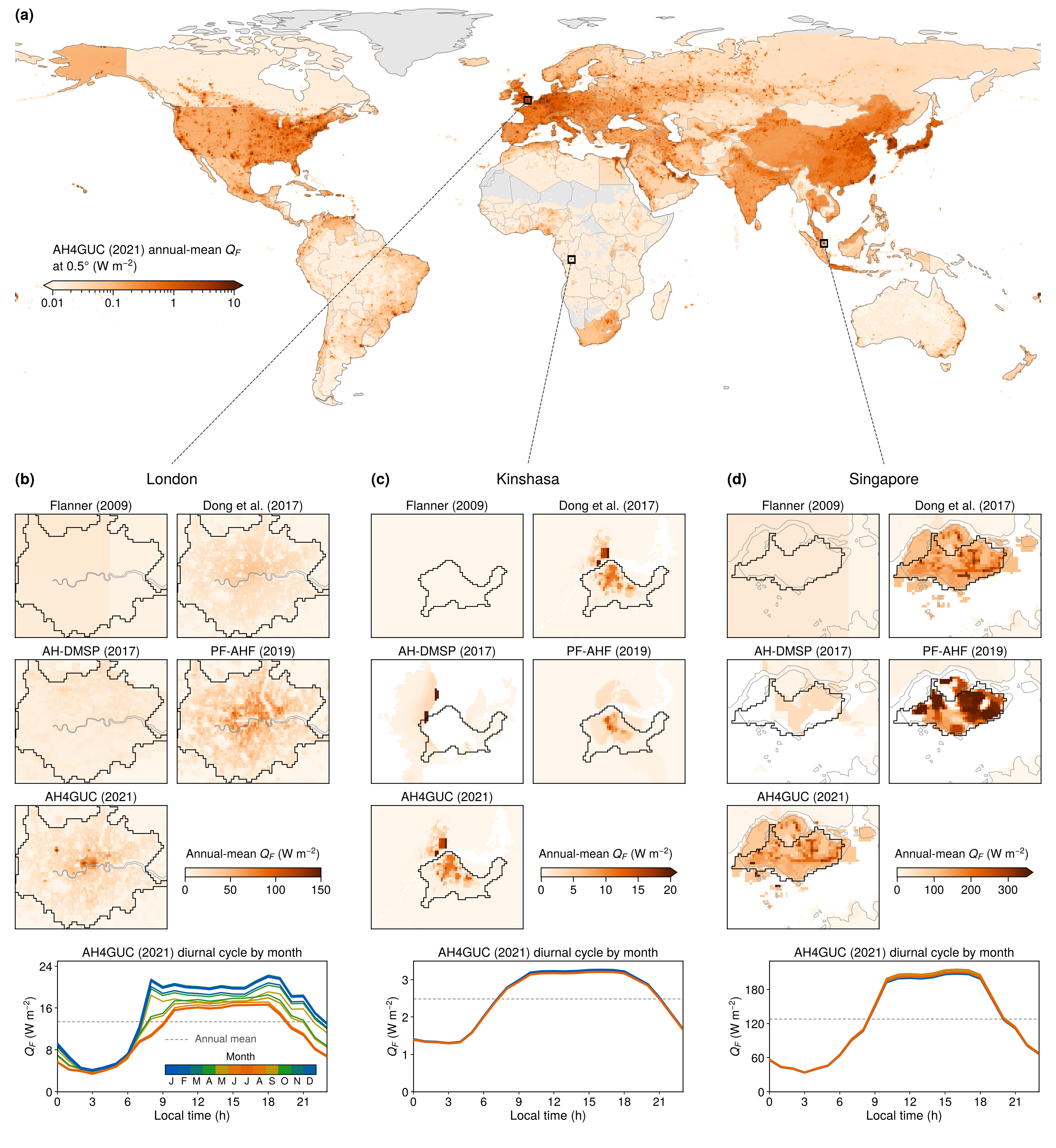}
    \caption{Comparison of global anthropogenic heat flux ($Q_{F}$) datasets. (a) Global distribution of annual mean $Q_{F}$ from \citeA{varquez_2021} at 0.5$^{\circ}$ resolution on a logarithmic scale (plate carr\'ee projection). (b)--(d) Annual mean $Q_F$ from five readily-accessible datasets (see Table~\ref{tab:global_qf_comparison}) for three cities representing distinct climatological and economic regimes: (b) London, (c) Kinshasa, and (d) Singapore, with a separate colorbar for each city. Black outlines mark the urban center of each city in the Global Human Settlement Urban Centre Database R2024A \cite{ghs_ucdb_2024}; thin gray lines mark the coastline (Eurostat GISCO, 1:1 million). Cells without anthropogenic heat are shown in white. The bottom row of (b)--(d) shows the diurnal cycle of $Q_F$ from \citeA{varquez_2021} averaged within the black outline of each urban center for each month; the dashed line marks the annual mean.}
    
    \label{fig:global_comparisons}
\end{figure}

Figure~\ref{fig:global_comparisons} illustrates these uncertainties by comparing several global datasets over Kinshasa, London, and Singapore. Kinshasa provides a useful test case because it is a rapidly growing tropical megacity with limited local urban climate observations (Figure~\ref{fig:global_comparisons}c). In such a data-sparse region, global datasets can provide a first approximation of anthropogenic heat forcing, but they also reveal substantial differences: the \citeA{flanner_2009} dataset produces a near-zero field over Kinshasa (and nearly uniform values in the other two cities) owing to its coarse resolution, whereas finer resolutions mainly disagree according to the treatment of spatial disaggregation: from relying mainly on nighttime lights products \cite{ah_dmsp_2017}, population datasets \cite{pf_ahf_2019} or a combination of both \cite{dong_2017,varquez_2021}. While verifications have been conducted for these datasets, through comparisons with either other $Q_{F}$ proxies such as nighttime lights \cite{pf_ahf_2019} or bottom-up estimates \cite{dong_2017,varquez_2021} including power-plant locations \cite{varquez_2021}, their spatial contrasts serve as a reminder that these global datasets are not direct observations of $Q_F$; they are empirical estimates that encode assumptions about where human activity, energy use, and anthropogenic heat are located.

The comparison for London  demonstrates that global datasets tend to agree more closely in their city-scale averages than in their representation of localized hotspots (Figure~\ref{fig:global_comparisons}b). Local top-down and bottom-up inventories estimate annual mean $Q_F$ values of approximately 5.79--10.9 W m$^{-2}$, with central London reaching 57.5--123.5 W m$^{-2}$ \cite{CapelTimms2020GMD,Iamarino_2011,harrison_1984}. Modeling studies suggest even higher peak daytime values exceeding 300 W m$^{-2}$ in WRF-SUEWS simulations \cite{suews_coupling_2024}. Averaged within the urban center of London (black outline in Figure~\ref{fig:global_comparisons}b), the global datasets give city-mean values of approximately 9--22 W m$^{-2}$, but they differ substantially in their representation of hotspots. %

For Singapore, the global datasets  not only differ in their representation of hotspots (Figure~\ref{fig:global_comparisons}d), but also depart from local estimates in the city-scale average. Local building inventories estimate a building-only city-wide mean of approximately 10.1 W m$^{-2}$ and a commercial peak of 663 W m$^{-2}$ in a single 200-m grid cell \cite{he_2023}. Neighborhood-scale inventories that also include traffic and human metabolism give peak hourly $Q_F$ of approximately 113 W m$^{-2}$ in a commercial district and 13--17 W m$^{-2}$ in residential areas \cite{quah_2012}. By contrast, averaged within the urban center (black outline in Figure~\ref{fig:global_comparisons}d), the \citeA{dong_2017}, AH4GUC, and PF-AHF datasets give approximately 125--210 W m$^{-2}$, exceeding even the 113 W m$^{-2}$ peak in the commercial district \cite{quah_2012}, whereas \citeA{flanner_2009} and AH-DMSP give approximately 27 and 23 W m$^{-2}$, respectively. Singapore thus illustrates the challenges that compact, densely built cities pose for global datasets. %

The global dataset comparison highlights two major sources of uncertainty related to spatial representation. The first is spatial smoothing: coarse grids dilute compact urban cores, transport corridors, and high-emitting industrial zones. The second is proxy sensitivity: assumptions about spatial allocation can produce substantially different $Q_F$ fields even when aggregate energy consumption is similar. For example, products that incorporate point-source information, such as Visible Infrared Imaging Radiometer Suite Nightfire, add local maxima at industrial sites, such as Jurong Island in Singapore, that are missing in other datasets \cite{varquez_2021}. Temporal representation is a further source of uncertainty. In AH4GUC (Figure~\ref{fig:global_comparisons}b--d, bottom row), heating demand keeps winter $Q_F$ above summer values throughout the day in London, whereas the diurnal cycle in Kinshasa and Singapore barely changes from month to month. These variations essentially follow the scheme designed by \citeA{dong_2017}, in which monthly $Q_F$ is scaled from its annual mean according to the monthly mean air temperature, using an empirical sensitivity of energy use to air temperature, and the diurnal cycle is taken from profiles derived for Tokyo \cite{moriwaki_2008}. These uncertainties do not undermine the usefulness of global datasets, which are indispensable for climate modeling and for regions lacking local inventories. However, they should be interpreted as scale-dependent, empirical forcing estimates rather than direct observations of $Q_F$.

The studies reviewed in this section suggest that our ability to characterize anthropogenic heat as an urban climate forcing has substantially advanced, revealing its strong spatial heterogeneity, temporal variability, sectoral breakdowns, and scale dependence. However, the magnitude and representation of the forcing alone do not determine its climatic significance. The same anthropogenic heat flux can produce markedly different temperature responses depending on the characteristics of the urban environment and the way $Q_F$ is introduced into the urban climate system. Quantifying these responses therefore requires moving beyond examining the forcing itself to investigating the sensitivity of urban temperatures to anthropogenic heat flux, which is the focus of the next section.

\section{Sensitivity} 
\label{sect:sensitivity}

Anthropogenic heat sensitivity quantifies the temperature response associated with a unit change in anthropogenic heat flux. Available estimates are derived from both observations and numerical simulations. Here, we classify a study as observational or numerical according to whether the temperature response is measured or simulated; an observational estimate may therefore still rely on model-derived estimates of anthropogenic heat flux.

Observational estimates remain relatively uncommon because isolating the temperature response to anthropogenic heat flux from concurrent variations in meteorological conditions, land cover, and other urban processes is inherently difficult. Studies that successfully constrain anthropogenic heat sensitivity using observations are therefore particularly valuable and are discussed first below, followed by the much larger body of numerical modeling studies. In numerical experiments, anthropogenic heat flux may either be prescribed or generated interactively by a coupled BEM or another model component. As discussed in Section~\ref{sect:theory}, prescribed-forcing experiments yield forcing-based sensitivity ($S_f$), whereas coupled simulations generally yield effective sensitivity ($S_e$), unless the total anthropogenic heat flux change can be separated into forcing and source-feedback contributions. Observational estimates generally face the same decomposition problem and are therefore more comparable to $S_e$ than to $S_f$.

\subsection{Observational studies}

Observational studies on the climatic effects of anthropogenic heat flux, or more broadly urbanization, generally exploit natural temporal or spatial variations in human activity, following the conceptual framework articulated by \citeA{lowry1977}. Temporal variations, such as weekday–weekend contrasts \cite{mitchell1953,kikegawa2014observed} or larger disruptions like G20 summit and COVID-19 lockdown \cite{pal2021,meng2023}, provide ``natural experiments’’ in which changes in anthropogenic activity induce corresponding changes in urban temperature. These ``natural experiments’’ have also been studied numerically \cite<e.g.,>{kikegawa2014observed,Nakajima2021UrbanClimate,Kikegawa2022AE,Takane2022NPJ}. Other studies instead examine spatial co-variations between anthropogenic activity indicators and urban temperatures or UHI intensities using dense station networks or remote sensing products \cite{he2020,jin_2020,zhu2017}. 

Two principal challenges arise when interpreting these studies. First, observational studies rarely quantify the anthropogenic heat flux change ($\Delta Q_F$) associated with the observed temperature response ($\Delta T$). Second, even when $\Delta Q_F$ is estimated, observed temperature variations reflect the combined influence of meteorological variability, land-cover heterogeneity, topography, and other urban processes. %
Consequently, observed correlations between $\Delta T$ and $\Delta Q_F$ do not necessarily reflect causation, and it is generally difficult to separate the contribution of $\Delta Q_F$ to the observed $\Delta T$ from that of other factors.

To address these challenges, particularly the second, studies have adopted statistical approaches designed to reduce the influence of co-varying processes, including stepwise multiple linear regression models \cite{chen2023}, partial least squares regression models \cite{chen2025}, linear mixed-effects models \cite{qian2023}, and machine-learning methods such as random forest \cite{chen2025b} and Light Gradient Boosting Machine \cite{sheng2025}. Nevertheless, interpreting the results using anthropogenic heat sensitivity remains challenging. First, the reported results are not always directly comparable to the anthropogenic heat sensitivity defined in the present review. For example, some studies report standardized regression coefficients \cite<e.g.,>{qian2023,chen2025b}, whereas others stratify the temperature response into different categories \cite<e.g.,>{sheng2025}, making it difficult to translate the results into physically interpretable sensitivities with units of K (W m$^{-2}$)$^{-1}$. Second, the inferred relationships sometimes appear physically counterintuitive. For example, some models yield negative coefficients during certain periods or for certain anthropogenic heat flux components  \cite<e.g.,>{chen2023,chen2025}, despite the expectation that anthropogenic heat flux should generally induce warming effects. Likewise, although anthropogenic heat flux effects are expected to be stronger at night and during winter, some studies identify anthropogenic heat flux variables only in summer prediction equations but not in winter equations \cite<e.g., see Table 2 of >{chen2023} or only for daytime but not for nighttime \cite<see Table 4 of>{chen2025}. These examples illustrate that anthropogenic heat sensitivities inferred statistically from observational data depend not only on the underlying physical relationship but also on the statistical model formulation, including the choice of predictors, variable transformations, spatial and temporal aggregation, and treatment of co-varying factors.

Another approach exploits both spatial and temporal contrasts in human activity through a difference-in-differences design. For example, \citeA{kikegawa2014observed} measured near-surface air temperatures in pairs of commercial and residential districts in Tokyo and Osaka, Japan, together with hourly district-level electricity demand, from which, with estimated motor-fuel consumption, they derived an energy-consumption-based anthropogenic heat flux. In the commercial districts, this flux decreased by about half from weekdays to weekends, whereas it remained small and nearly constant in the residential districts. On calm and clear days, the commercial--residential temperature difference decreased from weekdays to weekends in proportion to the reduction in anthropogenic heat flux in the commercial districts. Because the residential district serves as a control that removes background meteorological variability common to both districts, and because rooftop-level wind and solar radiation could not explain the decrease, the differential response was attributed to the reduction in anthropogenic heat flux in the commercial districts. This approach yielded afternoon sensitivities of approximately 0.01~K~(W~m$^{-2}$)$^{-1}$ in both cities. This estimate is notable because it falls within the range commonly reported in numerical modeling studies (Section~\ref{subsect:typical-magnitude}), thereby providing independent observational evidence consistent with model results. Similarly, \citeA{ohashi2016impact} examined seasonal variations in weekday–weekend temperature contrasts across multiple urban districts in Osaka, Japan, and related them to seasonal differences in electricity consumption. However, their temperature response and electricity-consumption change were evaluated over different time periods, making the resulting ratio difficult to interpret. This example motivates the temporal and spatial consistency criteria introduced in Section~\ref{subsect:numerical} for synthesizing anthropogenic heat sensitivities from numerical studies.

\subsection{Numerical studies}
\label{subsect:numerical}

Numerical studies of anthropogenic heat flux effects have evolved alongside the development of urban climate models. Early studies primarily used bulk surface energy balance models, either in offline mode or coupled to one-, two-, and three-dimensional atmospheric models. Beginning around the 2000s, urban canopy models (UCMs) introduced more explicit representations of urban geometry and canopy air energy exchange, and have since been coupled to weather, climate, and Earth system models. Other studies have also employed computational fluid dynamics tools to resolve urban flow and transport processes at building and street-canyon scales. Because anthropogenic heat sensitivities are rarely reported explicitly in these studies, this subsection focuses on those from which forcing-based or effective sensitivities can be extracted or reasonably inferred. Rather than attempting a comprehensive survey of the numerical modeling literature, we synthesize 42 studies across time\nocite{ghadban2020,cao2017,Nakajima2021UrbanClimate,lin2008urban,khanh2025impact,mccarthy2010,mussetti2022electric,huang2021,chen2016Sensitivity,Xie2024SCS,bueno2012resistance,karlicky2026,wang2023Sensitivity,Xin2023JGR,Yamaguchi2025UrbanClimate,Sun2026JAMES,bueno2011,gutman1975,chen2024modelling,gonzalezAparicio2014,sorbjan1982,hu2012,fan2005,ko2026modeling,yu1975,tao2022,feng2014,salvati2019,myrup_1969,lei2022,neunhauserer2007,li2013,aoyagi2012,swaid1990b,block_2004,atwater1972,Li2024Structural,wang2019,lyu2024,swaid1993,Kikegawa2022AE,ming2021}, from which we extract 79 sensitivity estimates. The 42 studies were identified from our literature corpus (Text~S1 in Supporting Information) as those from which a sensitivity could be diagnosed under the consistency criteria described below. Identifying the studies and diagnosing the values involved an iterative process combining large-language-model-assisted extraction with expert judgement, and each diagnosed value was verified against the original publication. Our synthesis focuses on local- to regional-scale estimates. It includes estimates from global models diagnosed over urban grid cells, but not estimates diagnosed from global averages, because too few studies yield the latter to support a synthesis; the challenges of diagnosing anthropogenic heat sensitivity from global averages are discussed in Section~4.2.3. %

To facilitate meaningful comparisons across models, scales, and study settings, anthropogenic heat sensitivity is diagnosed only when $\Delta T$ and $\Delta Q_F$ are averaged over consistent temporal periods and spatial domains. For example, a daily mean temperature response is paired with a daily mean anthropogenic heat flux, whereas a domain-mean temperature response at a given hour is paired with the domain-mean anthropogenic heat flux over the same hour. In contrast to previous work that included estimates based on the ratio of a temperature response to a peak or maximum daily anthropogenic heat flux \cite{Wang2023ERL,Xie2024SCS,Li2024NatClimate}, such estimates are excluded from our synthesis. We nevertheless recognize that different averaging operators could, in principle, be applied to $\Delta T$ and $\Delta Q_F$, and discuss the challenges associated with defining appropriate temporal and spatial averaging operators in Section~\ref{subsect:defining-sensitivity}. In addition, our synthesis focuses on near-surface air temperature such as the canopy air temperature and 2-m air temperature and does not consider land surface temperature.

\subsubsection{Typical magnitude and major sources of variability in anthropogenic heat sensitivity}
\label{subsect:typical-magnitude}

\begin{figure}[htbp]
    \centering
    \includegraphics[width=\textwidth]{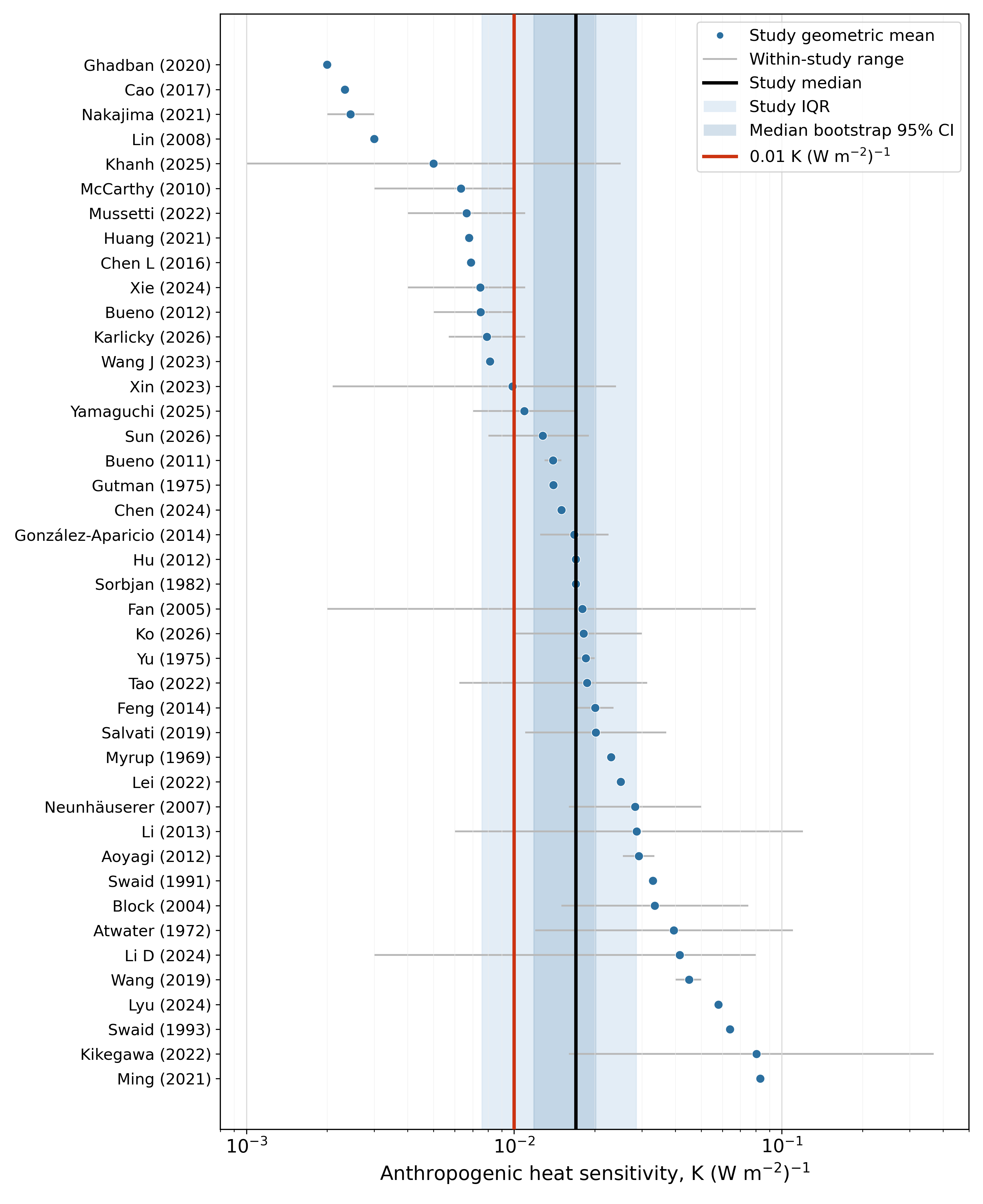}
\caption{A study-level synthesis of reported anthropogenic heat sensitivity K (W m$^{-2}$)$^{-1}$. Full reference information corresponding to each abbreviated study label is provided in Table S2. Each point represents the geometric mean diagnosed from one study, and, where present, a horizontal line shows the range (see Table S2). The vertical black line marks the study-level median, which is $0.017$ K (W m$^{-2}$)$^{-1}$, while shaded bands show the interquartile range (IQR) and bootstrap 95\% confidence interval for the median. Studies are ordered from lower to higher sensitivity. Study labels use only the first author for simplicity, with initials added where needed to distinguish duplicate labels. The vertical red line marks the rule-of-thumb value of 0.01 K (W m$^{-2}$)$^{-1}$.
}    \label{fig:sensitivity}
\end{figure}

\begin{figure}[htbp]
    \centering
    \includegraphics[width=\textwidth]{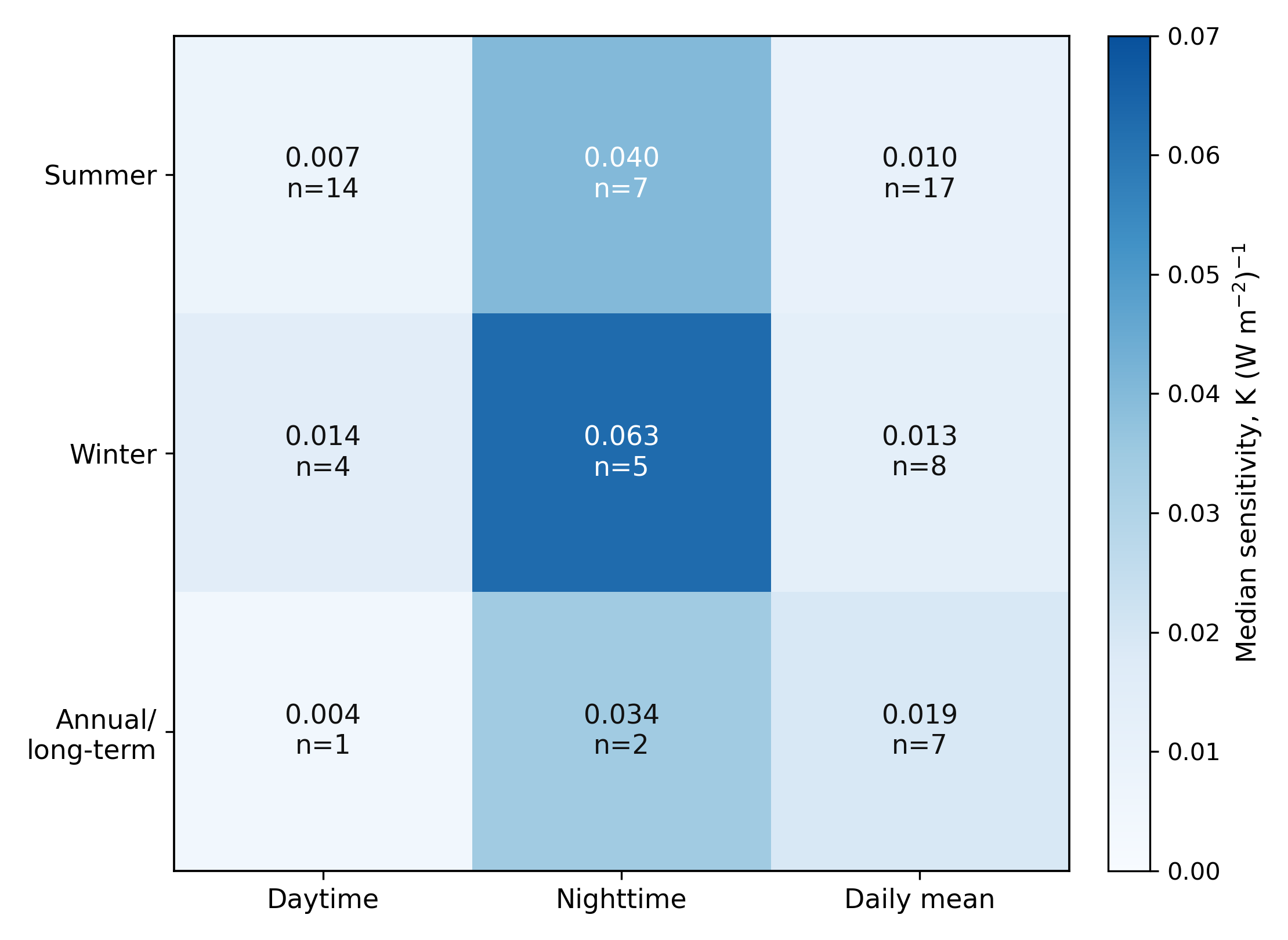}

\caption{Median anthropogenic heat sensitivity grouped by season and time of day. Numbers in each cell show the median sensitivity in K (W m$^{-2}$)$^{-1}$ and the number of estimates. Only estimates with classifiable season and time of day are included, so the counts do not sum to 79. The cell medians pool estimates across studies and are not calculated from matched day--night pairs.}
\label{fig:sensitivity2}
\end{figure}

\begin{figure}[htbp]
    \centering
    \includegraphics[width=\textwidth]{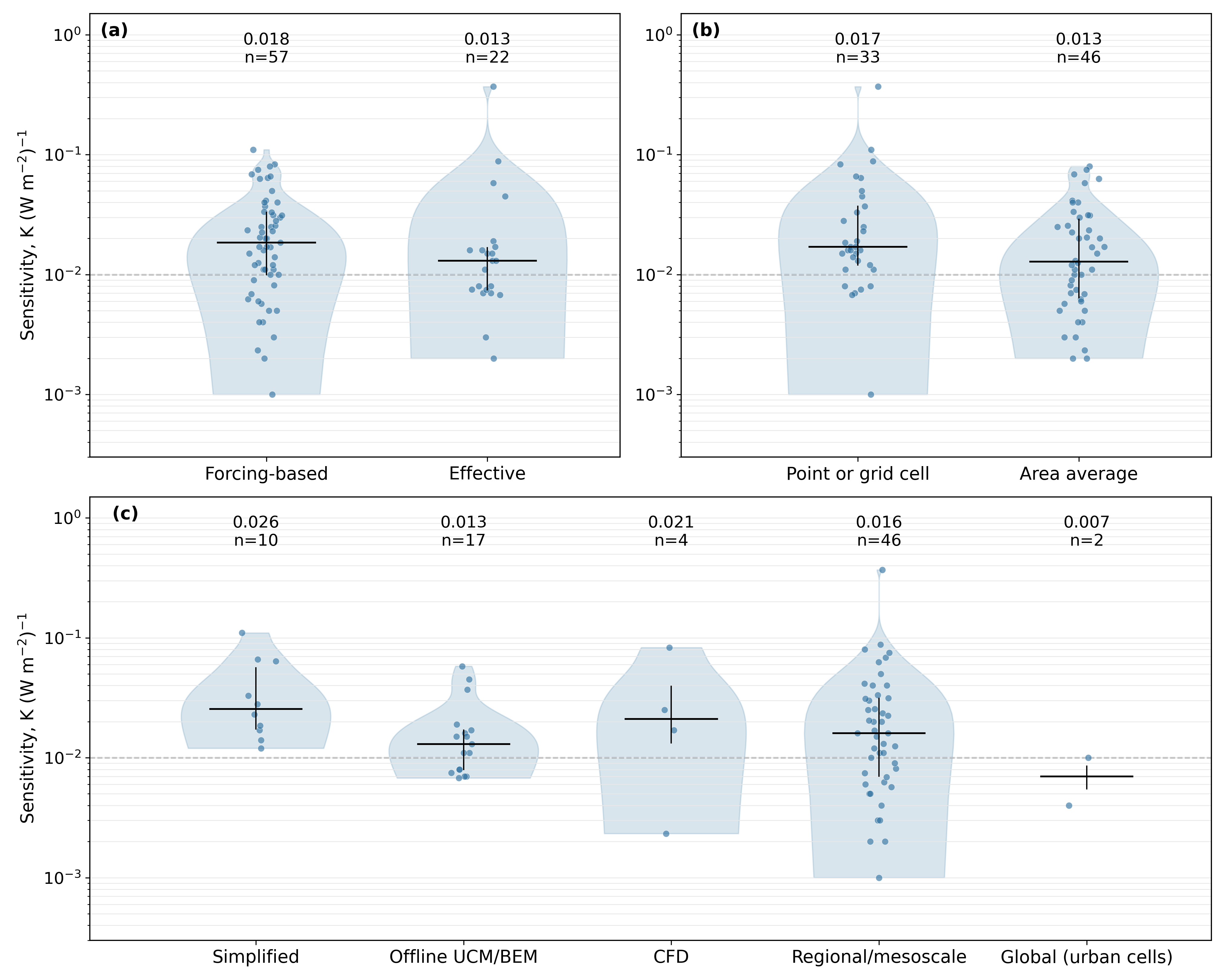}
\caption{Diagnosed anthropogenic heat sensitivities grouped by methodological factors: (a) sensitivity definition, (b) spatial representation, and (c) model type. Points show individual estimates ($n=79$), vertical lines show the interquartile range, and horizontal bars show the median; numbers above each group give the median of the estimates in K (W m$^{-2}$)$^{-1}$ and their number. In (b), slopes of regressions across grid cells are treated as area averages, whereas regressions over time at a single point or grid cell are treated as point estimates. Treating the cross-grid regression slopes as point estimates instead does not change the conclusion that the two groups do not differ significantly. The dashed line marks 0.01 K (W m$^{-2}$)$^{-1}$. Simplified models are slab models, some coupled to one- or two-dimensional boundary-layer models, and one-dimensional canopy-air models; UCM/BEM denotes offline urban canopy and building energy models; global models were diagnosed over urban grid cells.}
\label{fig:sensitivity3}
\end{figure}

A study-level synthesis of diagnosed sensitivities shows that reported values vary by more than an order of magnitude, reflecting differences in model formulations, study regions, temporal and spatial scales, and experimental design (Figure~\ref{fig:sensitivity}). The median study-level sensitivity is 0.017 K (W m$^{-2}$)$^{-1}$, with an interquartile range of 0.008--0.029 K (W m$^{-2}$)$^{-1}$. Given this variability, a value of order $0.01$ K (W m$^{-2}$)$^{-1}$ should be regarded as a typical magnitude or a rule-of-thumb value, rather than a consensus value, of the anthropogenic heat sensitivity of local- to regional-scale near-surface air temperature, which can be multiplied by a given $\Delta Q_F$ for a quick first estimate of the temperature response. Interestingly, this value is consistent with the observational analysis of \citeA{kikegawa2014observed}.

Across all the studies examined here, median sensitivities are larger at night than during the day in both summer and winter (Figure \ref{fig:sensitivity2}). In the extracted estimates, the median sensitivity increases from approximately 0.007 K (W m$^{-2}$)$^{-1}$ during summer daytime to 0.040 K (W m$^{-2}$)$^{-1}$ during summer nighttime, and from 0.014 K (W m$^{-2}$)$^{-1}$ during winter daytime to 0.063 K (W m$^{-2}$)$^{-1}$ during winter nighttime. This pattern is consistent with the physical understanding that anthropogenic heat flux warms the near-surface atmosphere more efficiently when turbulent mixing is weaker, boundary layers are shallower, and background radiative forcing is smaller.

The typical magnitude of anthropogenic heat sensitivity and its systematic day–night contrast are already evident in early modeling studies. In fact, the 7 pre-2000 modeling studies included in our synthesis yield a median sensitivity that is not statistically distinguishable from that of the remaining post-2000 studies based on a permutation test, and one of the earliest studies, \citeA{atwater1972}, had already reported both day–night and summer–winter contrasts. What could not be assessed from the pre-2000 literature, because all of those studies prescribed $Q_F$ as an external forcing, is whether the definition of anthropogenic heat sensitivity matters. In the present synthesis, forcing-based sensitivities ($S_f$) have a somewhat higher median than effective sensitivities ($S_e$), approximately 0.018 K (W m$^{-2}$)$^{-1}$ for $S_f$ compared with 0.013 K (W m$^{-2}$)$^{-1}$ for $S_e$ (Figure \ref{fig:sensitivity3}a), although the difference is not statistically significant.

At first glance, a lower $S_e$ may appear counterintuitive. Because positive source feedbacks, which are more commonly represented in the studies reporting $S_e$ here, amplify the temperature response to a given anthropogenic heat forcing, one might expect studies that include such feedbacks to exhibit higher, rather than lower, anthropogenic heat sensitivities. However, it is fully consistent with the definitions of $S_f$ and $S_e$. The forcing-based sensitivity $S_f$ relates the temperature response to the forcing alone, whereas the effective sensitivity $S_e$ is defined using the total anthropogenic heat flux change, which includes both the forcing and the additional anthropogenic heat generated through source feedback. The apparent contradiction can therefore be reconciled by recognizing that, although a positive source feedback amplifies the temperature response, it simultaneously increases the diagnosed $\Delta Q_F$ in the denominator of $S_e$. Consequently, a lower value of $S_e$ does not necessarily imply a weaker thermal response, but instead reflects the different definition of anthropogenic heat sensitivity.

More broadly, this distinction emphasizes that the intuitive connection between positive feedback and increased climate sensitivity applies only to the forcing-based sensitivity $S_f$ defined here (see Eqs. \ref{eq:sensitivity-with-feedback} and \ref{eq:feedback-parameters}), whereas the effective sensitivity $S_e$ cannot be interpreted using the same relationship. Accordingly, studies that represent source feedback but seek to interpret their sensitivity results from a feedback perspective, to be discussed in Section \ref{sect:feedback}, must separately define a forcing-based sensitivity. The practical challenges involved in defining such a forcing-based sensitivity are discussed further in Section~\ref{subsect:diagnosing-feedback}.

Two further factors that might be expected to affect the diagnosed sensitivity are the spatial representation of the estimate and the type of model used. The median sensitivity is 0.017 K (W m$^{-2}$)$^{-1}$ for estimates diagnosed at single points or individual grid cells and 0.013 K (W m$^{-2}$)$^{-1}$ for those diagnosed from area averages (Figure \ref{fig:sensitivity3}b). Across model types, the medians range from approximately 0.007 K (W m$^{-2}$)$^{-1}$ for global models diagnosed over urban grid cells to 0.026 K (W m$^{-2}$)$^{-1}$ for simplified models (Figure \ref{fig:sensitivity3}c), but the distributions overlap substantially. Neither factor yields statistically significant differences, whether tested across individual estimates or at the study level. These comparisons should be interpreted with caution, because some groups contain few estimates and estimates from the same study are not necessarily independent.

\subsubsection{Methodological sources of variability in anthropogenic heat sensitivity}

The synthesis above highlights that anthropogenic heat sensitivity varies most clearly with time of day. Beyond the broad categories examined above, more subtle sources of variability arise from how sensitivity is diagnosed.

The first issue concerns the definition of anthropogenic heat flux itself. Although $Q_F$ is broadly interpreted as the sensible heat released by human activities into the urban environment in this review, different studies have adopted different definitions \cite{Liu2022ACP, Xie2024SCS}. Traditional inventory-based approaches typically estimate $Q_F$ from energy consumption, whereas building energy studies may define it in terms of the net heat exchanged between buildings and the outdoor urban environment, including heat transported through ventilation, infiltration, and HVAC systems \cite{sailor_methods}. As emerging technologies such as air-source heat-pumps become more widespread, these distinctions are likely to become increasingly important. Consequently, reported anthropogenic heat sensitivities may differ simply because the underlying definition of $Q_F$ differs.

The second issue concerns the reference area used to define $Q_F$, which directly affects its reported magnitude \cite{best2015}. Establishing a consistent reference area is challenging because anthropogenic heat is represented differently across models. Mesoscale models, for example, often prescribe $Q_F$ as an areal flux density, whereas computational fluid dynamics simulations \cite<e.g.,>{zhang2024relative} and laboratory experiments \cite<e.g.,>{liu2025modeling} may prescribe anthropogenic heat as line or volumetric heat sources. Comparisons with conventional area-based definitions of $Q_F$ are therefore less straightforward.

A related issue is that, for the diagnosed sensitivity to be physically meaningful, the reported temperature and $Q_F$ should correspond to compatible spatial supports, such as a street canyon, urban tile, model grid cell, or observation footprint. In many models, however, they do not. For example, in the Single-Layer Urban Canopy Model (SLUCM) coupled to the Weather Research and Forecasting (WRF) model, $Q_F$ is expressed per unit impervious urban land area, including both roofs and street canyon floors, whereas canopy air temperature represents only the air within the street canyon. The diagnosed sensitivity of canopy air temperature to $Q_F$ therefore depends strongly on the ratio of roof area to canyon-floor area \cite{Li2024Structural}. Similarly, 2-m air temperature represents the entire model grid cell, whereas $Q_F$ is applied only over the impervious urban fraction of that grid cell. The diagnosed sensitivity based on 2-m air temperature therefore depends on the impervious urban land fraction \cite{Li2024Structural}.

The third issue concerns how sensitivity is computed. Because $\Delta T$ and $\Delta Q_F$ are not always available over consistent spatial domains and temporal periods, estimating their ratio may require judgment in pairing the two quantities, producing a range of plausible sensitivity estimates that reflects this ambiguity. The choice of spatial and temporal averaging operators, which varies across studies in Figure \ref{fig:sensitivity}, also plays an important role. %
Other studies estimate sensitivity by fitting a linear relationship between $\Delta T$ and $\Delta Q_F$ and interpreting the fitted slope as the sensitivity. This approach is generally less sensitive to individual outliers. However, the fitted sensitivity depends on the number, range, and distribution of the sampled data points, which are themselves influenced by the chosen spatial and temporal averaging operators. The effects of these averaging operators are discussed further in Section~\ref{subsect:defining-sensitivity}.

\subsubsection{Sources of uncertainty in anthropogenic heat sensitivity}

The previous subsection focuses on methodological choices that can lead to different reported anthropogenic heat sensitivities. Here, we distinguish these methodological differences from uncertainty in the diagnosed sensitivity itself, which arises primarily from internal variability and structural uncertainty.

Because the climate system is chaotic, the simulated temperature response to anthropogenic heat flux contains both the forced signal and internally generated fluctuations. This issue is particularly important in global simulations, where the globally averaged $Q_F$ is small. The treatment of this uncertainty has evolved considerably over time. \citeA{washington_1972} found that the simulated response to anthropogenic heat flux was comparable to the model’s internal variability. Subsequent studies, including \citeA{flanner_2009} and \citeA{chen2014anthropogenic,chen_2016}, diagnosed the response from a single control run without $Q_F$ and a perturbation run with $Q_F$. More recent studies have increasingly adopted ensemble approaches. \citeA{jacobson2014effects} characterized internal variability using multiple control runs, whereas \citeA{zhang2013energy} and \citeA{chen2019seasonal,chen2023anthropogenic} performed multiple control/perturbation runs initialized from different climate states to estimate the significance of ensemble-mean response. However, the published results do not always provide sufficient information to quantify uncertainty in the diagnosed sensitivity. %
Future studies should therefore quantify anthropogenic heat sensitivity using ensemble simulations and report both the ensemble-mean sensitivity and its statistical uncertainty, particularly when the response is averaged globally, where the small magnitude of the globally averaged $Q_F$ makes the forced response difficult to separate from internal variability.

Another source of uncertainty is structural uncertainty associated with model formulation and the choice of physical parameterizations. Different urban climate models may represent land-surface processes, land-atmosphere interactions, and the pathways through which anthropogenic heat enters the urban climate system differently, leading to different temperature responses even when the total anthropogenic heat flux is the same. For example, \citeA{Li2024Structural} showed that variations in $Q_F$ release pathways and UCM structures can produce differences in anthropogenic heat sensitivity approaching an order of magnitude. In offline simulations, structural uncertainty can also arise from differences in the driving atmospheric forcing. For example, \citeA{Li2024NatClimate} forced machine-learning emulators of a physically based urban climate–BEM with projections from a multimodel ensemble of Coupled Model Intercomparison Project Phase 6 models, thereby quantifying the spread in future anthropogenic heat flux projections associated with intermodel differences in the simulated background climate. Such structural uncertainties are best assessed through coordinated comparisons across alternative urban model configurations and driving climate models, using common forcing magnitudes and standardized sensitivity diagnostics.

\subsection{Summary}

In summary, diagnosed anthropogenic heat sensitivities vary by more than an order of magnitude across studies, with a typical magnitude of order 0.01 K (W m$^{-2}$)$^{-1}$ for local- to regional-scale near-surface air temperature. Sensitivity tends to be larger at night than during the day, whereas differences associated with how it is defined, the spatial representation, and the model type are not statistically significant across the studies synthesized here, even though the distinction between forcing-based and effective sensitivities remains important for interpretation. At the same time, observational constraints remain sparse, and existing numerical estimates are affected by inconsistent definitions, methodological differences, internal variability, and structural uncertainty. Consequently, major research gaps remain in how anthropogenic heat sensitivity should be defined, diagnosed, interpreted, and compared across different models, locations, and spatial and temporal scales.

\section{Feedback} \label{sect:feedback}

As alluded to in Section \ref{sect:sensitivity}, forcing-based sensitivity ($S_f$) can be interpreted from a feedback perspective using Eqs. \ref{eq:sensitivity-with-feedback} and \ref{eq:feedback-parameters}. The interquartile range for $S_f$ in our synthesis is 0.01--0.033 K (W m$^{-2}$)$^{-1}$, corresponding to a total feedback parameter $\Lambda$ of approximately $-100$ to $-30$ W m$^{-2}$ K$^{-1}$. This total feedback parameter reflects the combined contributions of baseline, restoring, and source feedbacks, which are examined in this section. Here, we focus primarily on the small number of studies that explicitly quantified these feedback processes using feedback parameters or equivalent gain factors. Studies that described feedback-like behaviors associated with anthropogenic heat flux are discussed only briefly.

\subsection{Baseline and restoring feedbacks}

The underlying physics of baseline and restoring feedbacks has long been implicit in previous work. For example, many studies have recognized that anthropogenic heat sensitivity depends on boundary-layer depth, with weaker temperature responses when the boundary layer is deeper or the effective heating volume is larger \cite<e.g.,>{ohashi2016impact,khanh2024multi,khanh2025impact}. Although boundary-layer depth does not appear explicitly in our framework, its influence can still be interpreted from the perspective of the urban canopy layer, because a deeper boundary layer is often associated with more efficient heat exchange between canopy air and the overlying atmosphere, represented by a larger $C_{ha}$. Therefore, the weaker temperature response observed under deeper boundary layers is consistent with a smaller reference sensitivity and a stronger baseline restoring  (i.e., a larger absolute baseline feedback parameter) associated with larger $C_{ha}$ in our framework (Eqs. \ref{eq:S0} and \ref{eq:lambda conductances}). If anthropogenic heat flux further modifies boundary-layer depth (or $C_{ha}$), the resulting adjustment constitutes a restoring feedback.

\citeA{Wang2023ERL} provides, to our knowledge, the first explicit theoretical formalization of baseline and restoring feedbacks, which paves the way for the forcing--response--feedback framework developed in this review. The key difference is that \citeA{Wang2023ERL} did not include source feedback. Moreover, as noted in Section~\ref{subsect:interpreting-framework}, their decomposition of the total restoring feedback was expressed in terms of feedback parameters associated with heat conductances ($C_{ha}$ and $C_{hs}$) and temperatures ($T_s$ and $T_{atm}$), rather than in terms of $\lambda’$ and $c’$. 

Unlike source feedback, which can in principle be disabled by prescribing anthropogenic heat flux and preventing it from responding to temperature, baseline and restoring feedbacks are more difficult to isolate because they are internally realized within the urban climate system. The simulation design in \citeA{Wang2023ERL} was deliberately idealized to isolate a limited set of physical processes. Because they used offline simulations with prescribed  overlying atmospheric states, atmospheric adjustments to anthropogenic heat forcing could not feed back onto the urban canopy air energy budget. Their analysis therefore excluded  restoring feedbacks associated with $T_{atm}$ and focused on the baseline and restoring feedbacks associated with $C_{ha}$, $C_{hs}$, and $T_s$ (see Figure~\ref{fig:theory}).

\begin{figure}[htbp]
    \centering
    \includegraphics[width=\textwidth]{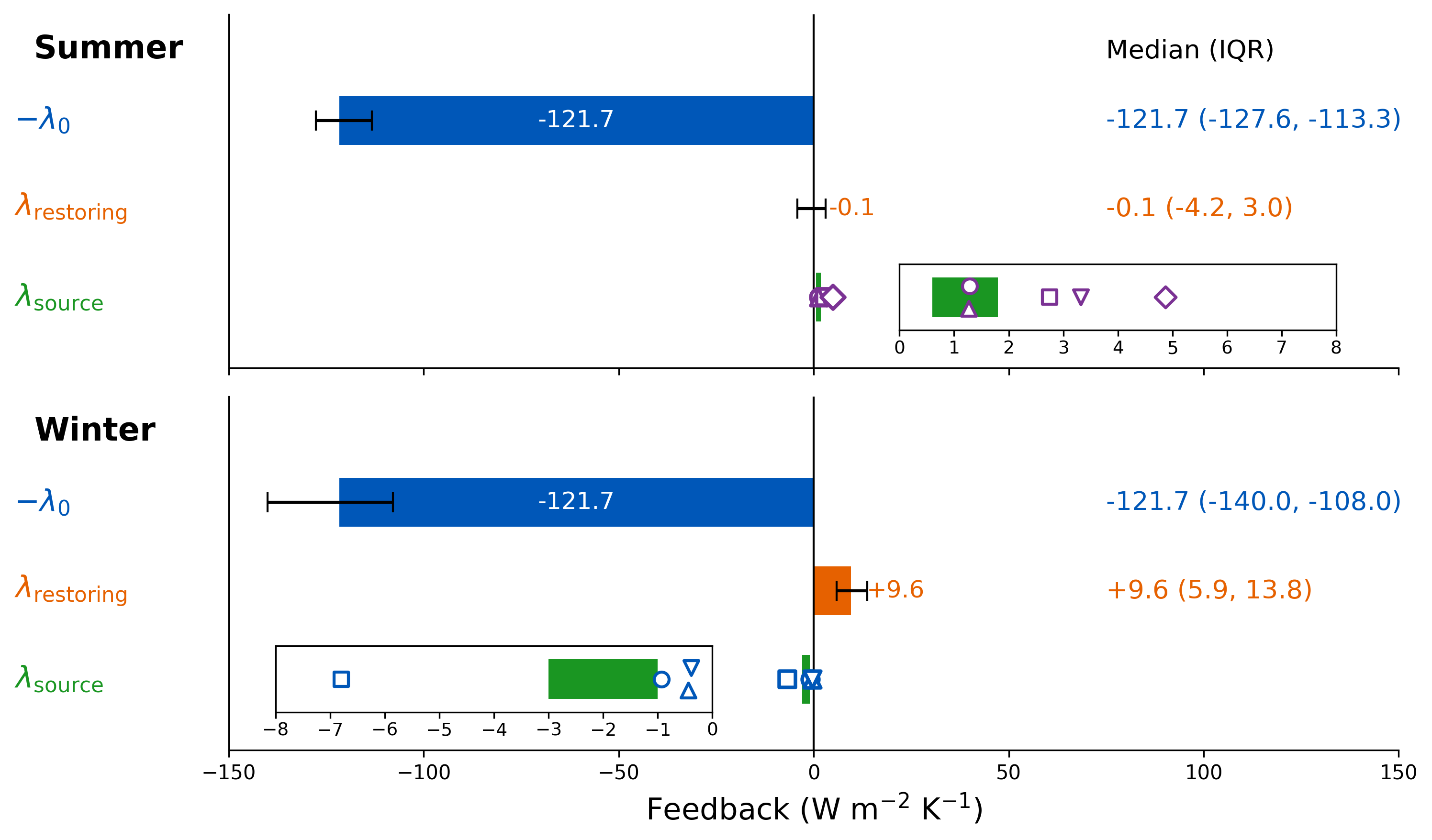}
\caption{
Comparison of the baseline feedback parameter ($-\lambda_0$), restoring feedback parameter ($\lambda_{\rm restoring}$), and source feedback parameter ($\lambda_\text{source}$). Positive values indicate amplifying feedbacks, whereas negative values indicate damping feedbacks. Blue and orange bars show domain-median values across urban grid cells within the CONUS in the study by \citeA{Wang2023ERL}, with horizontal error bars denoting the interquartile range (IQR, 25th--75th percentile). The source feedback parameter is represented using operational estimates drawn from the broader literature. Consequently, it should not be interpreted as directly comparable to the baseline and restoring feedback parameters. The three parameters are shown together solely for the purpose of qualitatively synthesizing their relative magnitudes and signs. The boxes on the $\lambda_\text{source}$ rows show zoomed-in views of the source feedback parameter, in which symbols for nearly identical values are offset vertically for legibility only. For the source feedback parameter, the green bars show the rule-of-thumb ranges from \citeA{GinzburgDemchenko2017}: 0.6 to 1.8 W m$^{-2}$ K$^{-1}$ for summer and $-3$ to $-1$ W m$^{-2}$ K$^{-1}$ for winter. In the summer panel, the purple symbols denote estimates derived from building energy model simulations: the circle and diamond represent the daytime estimates over residential and commercial areas from \citeA{Kikegawa2022AE}, with values of 1.29 (the average of 1.15 and 1.42) and 4.87 W m$^{-2}$ K$^{-1}$, respectively; the upward- and downward-pointing triangles represent office and residential areas from \citeA{Takane2019NPJ}, with values of 1.27 and 3.32 W m$^{-2}$ K$^{-1}$, respectively; and the square represents the residential estimate of 2.75 W m$^{-2}$ K$^{-1}$ derived from \citeA{bueno2012resistance}. In the winter panel, the blue symbols denote estimates derived from empirical energy demand--temperature relationships: the square represents Toulouse, with a value of $-6.8$ W m$^{-2}$ K$^{-1}$ \cite{pigeon_2007}; the circle represents Moscow, with a value of $-0.93$ W m$^{-2}$ K$^{-1}$ \cite{GinzburgDemchenko2017}; and the upward- and downward-pointing triangles represent the two reported estimates for St. Petersburg, with values of $-0.445$ and $-0.385$ W m$^{-2}$ K$^{-1}$, respectively \cite{ginzburg2021dependence}. 
}    \label{fig:feedback}
\end{figure}

The results of \citeA{Wang2023ERL} are highlighted in Figure~\ref{fig:feedback}, which shows the domain-median values of the baseline feedback parameter ($-\lambda_0$) and restoring feedback parameter ($\lambda_\text{restoring}$) across urban grid cells within the contiguous United States (CONUS), with horizontal error bars denoting the interquartile range (IQR). The baseline feedback parameter was computed analytically using Eqs.~\ref{eq:S0} and \ref{eq:lambda conductances}. The restoring feedback parameter was diagnosed from paired simulations without and with $Q_F$ using Eq.~\ref{eq:restoring-regrouped}, with the derivatives approximated by the simulated changes in heat conductances and surface temperature relative to the change in canopy air temperature. These feedback parameters were diagnosed at hourly scales and subsequently averaged across both day and night over 1980–1999 to provide a long-term climatological summary.

A key contribution of \citeA{Wang2023ERL} is that their results explain why many previous studies reported anthropogenic heat sensitivities of approximately 0.01 K (W m$^{-2}$)$^{-1}$. As shown in Figure~\ref{fig:feedback}, the domain-median baseline feedback parameter is approximately $-120$ W m$^{-2}$ K$^{-1}$ in both summer and winter. Consequently, the reference sensitivity is close to 0.01 K (W m$^{-2}$)$^{-1}$. Moreover, because the baseline feedback parameter is nearly an order of magnitude larger than the restoring feedback parameter, the baseline restoring process largely determines the magnitude of anthropogenic heat sensitivity.

While the domain-median baseline feedback parameter remains nearly unchanged between seasons, the restoring feedback parameter exhibits a much stronger seasonal dependence. The median restoring feedback parameter increases from approximately 0 W m$^{-2}$ K$^{-1}$ in summer to about 10 W m$^{-2}$ K$^{-1}$ in winter, producing a corresponding increase in anthropogenic heat sensitivity. The spatial variability, on the other hand, is controlled more by the baseline feedback parameter ($-\lambda_0$) than by the restoring feedback parameter ($\lambda_\text{restoring}$), as indicated by the much larger interquartile ranges. Because \citeA{Wang2023ERL} prescribed the anthropogenic heat flux perturbation uniformly, the spatial variability of temperature response cannot be attributed to spatial differences in the imposed forcing. Instead, it reflects spatial differences in the baseline and restoring feedbacks. In particular, they found that regions with smaller $\lambda_0$ exhibit stronger temperature responses to a given anthropogenic heat flux perturbation, whereas regions with larger $\lambda_0$ exhibit weaker responses.

More recently, \citeA{Li2024Structural} extended this analysis to multiple UCM configurations using land–atmosphere coupled simulations, thereby allowing atmospheric adjustment to contribute an additional restoring feedback. Although they did not use the exact terminology adopted here, their comparison across model configurations showed that both the reference sensitivity (or the baseline feedback parameter) and the strength of restoring feedbacks depend strongly on model structure, reflecting substantial structural uncertainty. %

Another restoring mechanism not explicitly represented in our canopy-air framework is longwave radiative cooling. Recent work by \citeA{khanh2025impact} showed that, when the urban boundary layer is treated as a well-mixed control volume and longwave radiation is assumed to be the only explicit heat-loss mechanism, accounting for radiative cooling improves simple models of anthropogenic heat impacts. As derived in Section~\ref{subsect:interpreting-framework}, the corresponding baseline feedback parameter is $-4\varepsilon\sigma T_0^3$, which is on the order of $-6$ to $-5$ W m$^{-2}$ K$^{-1}$ under typical atmospheric conditions.

\subsection{Source feedback}

The investigation of source feedback has been enabled primarily by coupling BEMs with urban land surface models, either driven by prescribed atmospheric forcing or coupled with atmospheric models. Examples include Multi-Layer Urban Canopy Model (CM)-BEM \cite{Kikegawa2003AE}, Building Effect Parameterization (BEP)-BEM \cite{Salamanca2010TAC}, Town Energy Balance (TEB)-BEM \cite{Bueno2012GMD}, Urban Tethys \& Chloris (UT\&C)-BEM \cite{Meili2025JAMES}, SLUCM-BEM \cite{Takane2024GMD}, and SLUCM-EnergyPlus \cite{Vahmani2022BuildEnv}. These coupled modeling tools have been widely used to examine interactions among urban microclimate, building energy use, and anthropogenic heat fluxes at city scales \cite{Ohashi2007JAMC,deMunck2013IJC,Salamanca2014JGR,Takane2017IJC,Xu2018JGR,Takane2019NPJ,Takane2022NPJ,Vahmani2022BuildEnv}. More recently, global models have begun to connect climate change, cooling and heating energy demand, and anthropogenic heat emissions in cities \cite{Li2024ACAdoption,Li2024NatClimate}; parameterizations have been developed to link transportation-related anthropogenic heat fluxes with urban climate \cite{Sun2026JAMES}; and attention has expanded to emerging energy-system transitions such as the deployment of air-source heat pumps, the broader electrification of building heating, and heat pump water heaters \cite{Meyer2024NatCommun,Takane2024GMD,Xie2024SCS,Yamaguchi2025UrbanClimate}.

These studies provide the conceptual and modeling basis for treating anthropogenic heat fluxes as temperature dependent and many have expressed this dependence in terms of energy demand amplification or temperature sensitivity of energy demand \cite{KondoKikegawa2003,Kikegawa2006AE,krpo2010impact, Salamanca2012IJC,Takane2019NPJ,Nakajima2022EnergyBuildings,Takane2023SCS,Takane2024GMD}. Although these studies generally do not characterize source feedback using explicit feedback parameters or gain factors, their reported energy demand sensitivities can, in principle, be translated into the source feedback parameter $\lambda_{\mathrm{source}}$, as shown below, provided the underlying assumptions are carefully considered. Some studies also quantify associated CO$_2$ emissions \cite<e.g.,>{Yamaguchi2025UrbanClimate}, but this pathway is distinct from the source feedback considered here, which describes the local feedback between urban temperature, energy demand, and anthropogenic heat fluxes. By contrast, the pathway through which rising urban temperatures increase CO$_2$ emissions, which in turn affect the global climate through radiative forcing, operates over much larger spatial and temporal scales and is therefore not considered here.

\citeA{GinzburgDemchenko2017} was among the first to frame the feedback between urban temperature, energy consumption, and anthropogenic heat flux using the concept of a gain factor ($g_A$). In the notation of the present review, this gain factor can be written as $g_A \equiv \lambda_\text{source} S_0 \equiv \lambda_\text{source}/\lambda_0$. Unlike the framework developed here, which focuses on canopy air temperature, \citeA{GinzburgDemchenko2017} developed a similar framework for surface temperature by introducing anthropogenic heat flux into the urban surface energy balance equation (Eq.~\ref{eq:SEB}). Although their framework focused on surface temperature and did not explicitly include restoring feedbacks, their estimates of $\lambda_0$ and $\lambda_\text{source}$ remain useful for understanding the importance of source feedback.

Neglecting ground heat flux and assuming a very large Bowen ratio so that latent heat flux can also be neglected, \citeA{GinzburgDemchenko2017} estimated a baseline feedback parameter ($-\lambda_0$) of order $-100$ to $-10$ W m$^{-2}$ K$^{-1}$ under what they called the maximum-advection limit. In this limit, local surface energy perturbations are rapidly removed by radiative and turbulent sensible heat exchange before substantially modifying the overlying atmosphere. In the minimum-advection limit, where the atmosphere is represented as a single equilibrating layer, the baseline feedback parameter ($-\lambda_0$) becomes much smaller, approximately $-1.8$ W m$^{-2}$ K$^{-1}$. \citeA{DemchenkoGinzburg2018} later showed that these limiting regimes represent opposite ends of a spectrum controlled by the horizontal length scale of anthropogenic heat forcing relative to the atmospheric thermal adjustment length. The maximum-advection estimate is broadly consistent with the domain-median baseline feedback parameter of approximately $-120$ W m$^{-2}$ K$^{-1}$ obtained by \citeA{Wang2023ERL}, which is notable because the two studies differ substantially, with one focusing on surface temperature in an idealized analytical model and the other on canopy air temperature in a UCM. %

\citeA{GinzburgDemchenko2017} then estimated the source feedback parameter $\lambda_\text{source}$ from empirical energy demand--temperature relations. For the heating seasons in Moscow, they obtained $\lambda_\text{source} \approx -0.93$ W m$^{-2}$ K$^{-1}$, where the negative sign indicates that warmer conditions reduce heating demand. Following the same approach, $\lambda_\text{source}$ can be estimated from other published work, for example approximately $-6.8$ W m$^{-2}$ K$^{-1}$ in Toulouse, France \cite{pigeon_2007} and $-0.445$ to $-0.385$ W m$^{-2}$ K$^{-1}$ in St. Petersburg, Russia \cite{ginzburg2021dependence}. Such estimates, however, require caution because the regression slope of an empirical energy demand--temperature relation does not necessarily correspond to the partial derivative defining $\lambda_\text{source}$. This distinction is discussed further in Section~\ref{subsect:diagnosing-feedback}.

An alternative approach is to use BEM simulations \cite<e.g.,>{bueno2012resistance,Takane2019NPJ,Li2024NatClimate} rather than empirical energy demand--temperature relations. In these models, building energy demand and the associated anthropogenic heat fluxes respond to evolving meteorological and climate conditions, allowing the temperature response of $Q_F$ to be diagnosed directly from simulations. For example, \citeA{bueno2012resistance} found that residential cooling energy demand increased by approximately 5\% per 1~K increase in maximum nighttime UHI intensity. Combined with their simulated $Q_F$ of approximately 55~W~m$^{-2}$ in residential area, this implies a source feedback parameter of about 2.75~W~m$^{-2}$~K$^{-1}$, assuming that anthropogenic heat flux scales proportionally with cooling energy demand. Similarly, \citeA{Takane2019NPJ} and \citeA{Li2024NatClimate} conducted simulations under future warming scenarios, although at vastly different spatial scales. For Osaka at 14:00 local time in August, the diagnosed source feedback parameters were 1.27 and 3.32~W~m$^{-2}$~K$^{-1}$ for office and residential areas, respectively, with the office estimate including only the sensible component \cite{Takane2019NPJ}. At the global annual mean scale, the reported source feedback parameters were approximately 0.02--0.05~W~m$^{-2}$~K$^{-1}$ for cooling and $-0.4$ to $-0.3$~W~m$^{-2}$~K$^{-1}$ for heating, which depend strongly on the assumed Shared Socioeconomic Pathway \cite{Li2024NatClimate}. However, because temperature changes in many such simulations are accompanied by changes in other meteorological variables, the diagnosed source feedback parameters may not correspond to strict partial derivatives that define $\lambda_{source}$.

As a general rule of thumb, \citeA{GinzburgDemchenko2017} combined representative temperature sensitivities of energy demand, approximately 5\% K$^{-1}$ for heating and 3\% K$^{-1}$ for cooling, with typical anthropogenic heat fluxes for London (20 W m$^{-2}$) and Tokyo (60 W m$^{-2}$) \cite{lucy_2011}, yielding estimated source feedback parameters of roughly $\lambda_\text{source} \sim 0.6$–1.8 W m$^{-2}$ K$^{-1}$ during warm seasons and $\lambda_\text{source} \sim -3$ to $-1$ W m$^{-2}$ K$^{-1}$ during cold seasons (Figure \ref{fig:feedback}). The assumed temperature sensitivities of energy demand are broadly consistent with the literature values as reviewed elsewhere \cite{santamouris2015impact}.
Combining these source feedback parameters with $\lambda_0$ shows why source feedback can become important under certain conditions. For example, if $\lambda_0 \sim 10$ W m$^{-2}$ K$^{-1}$ under weak-ventilation conditions, then $\lambda_\text{source} \sim 1.8$ W m$^{-2}$ K$^{-1}$ implies $g_A=\lambda_\text{source}/\lambda_0 \sim 0.18$, corresponding to an amplification of the reference temperature response by about 20\%.

Building on this gain-factor perspective, \citeA{Kikegawa2022AE} were among the first to apply it to analyzing results from coupled UCM-BEM simulations, using Osaka as a case study. They separately estimated $\lambda_{source}$ and $\lambda_0$ (or $S_0$), the two quantities that determine the gain factor. They used two energy-conserving simulations: a weekdays-run, in which all days were treated as weekdays in urban energy-consumption schedules, and a holidays-run, in which holiday or weekend schedules were imposed. The source feedback parameter $\lambda_\text{source}$ was estimated from the regression of $Q_F$ against $T$ in the weekdays-run, while $S_0$ was estimated from the regression of $\Delta T$ against $\Delta Q_F$, where $\Delta$ was taken as the difference between the weekdays-run and the holidays-run. Again, the extent to which these regression-based diagnostics correspond to the formal feedback parameters is discussed further in Section~\ref{subsect:diagnosing-feedback}.

\citeA{Kikegawa2022AE} illustrated the estimates for three Osaka districts in their Figure 15. The commercial district had a much larger source feedback parameter, with $\lambda_\text{source}=6.137$ W m$^{-2}$ K$^{-1}$, but also a larger  $\lambda_0=62.5$ W m$^{-2}$ K$^{-1}$. By contrast, the residential examples had smaller $\lambda_\text{source}$ and $\lambda_0$, especially the R7 district, where $\lambda_\text{source}=0.787$ W m$^{-2}$ K$^{-1}$ and $\lambda_0=2.71$ W m$^{-2}$ K$^{-1}$. When averaged over the daytime period and across urban grids, \citeA{Kikegawa2022AE} reported $\lambda_\text{source}$ of 4.87 W m$^{-2}$ K$^{-1}$ for commercial areas and 1.15–1.42 W m$^{-2}$ K$^{-1}$ for residential areas. The corresponding $\lambda_0$ were approximately 50 W m$^{-2}$ K$^{-1}$ for commercial grids and 6.25–7.14 W m$^{-2}$ K$^{-1}$ for residential grids. These two quantities imply gain factors of about 10\% in commercial areas and 20\% in residential areas.

\subsection{Summary}

Taken together, these pioneering studies suggest an emerging picture of anthropogenic heat feedback. The available estimates indicate that both source and restoring feedback parameters are generally of order 10 W m$^{-2}$ K$^{-1}$ or smaller, while estimates of the baseline feedback parameter span a much wider range across the available studies. In the long-term climatological estimates of \citeA{Wang2023ERL}, the baseline feedback parameter is of order 100 W m$^{-2}$ K$^{-1}$, comparable to the upper end of the maximum-advection range estimated by \citeA{GinzburgDemchenko2017}. Substantially smaller baseline feedback parameters have also been reported, for example, by \citeA{Kikegawa2022AE} for a residential district at 16:00 local time in summer, suggesting that the diagnosed baseline feedback parameter depends strongly on the temporal and spatial scales considered. 

These findings should, however, be regarded as preliminary, given the small number of studies and their differences in models, assumptions, diagnostic methods, and, importantly, the temporal and spatial scales over which the feedback parameters are diagnosed. Establishing whether these patterns generalize across cities, climates, and modeling systems requires more systematic investigations.

\section{Future outlook}
\label{sect:outlook}

The literature synthesized in this review reveals a clear imbalance: substantial progress has been made in quantifying the forcing and documenting its effects in applied contexts, but less generalizable understanding has emerged regarding the urban temperature response to that forcing and the feedbacks that amplify or damp the response. This imbalance is reflected in the textbook by \citeA{oke2017}, which devotes substantial attention to the magnitude, sources, and estimation of anthropogenic heat flux, but only a brief discussion of its climatic significance. This disparity does not necessarily imply that the climatic effects of anthropogenic heat flux have received little attention; rather, it suggests that the large body of observational and modeling work has primarily addressed specific cities, applications, and modeling systems, making it difficult to extract broader physical understanding.

Interestingly, however, a sensitivity concept was hinted at in the discussion of \citeA{oke2017}, who noted that “the atmospheric thermal response (increase in air temperature per unit energy flux density) depends on the whole surface energy balance \ldots{} and the mixing efficiency of the air in the roughness sublayer.” In retrospect, this statement foreshadowed both the sensitivity concept and the role of $C_{ha}$ and $C_{hs}$ in controlling the sensitivity. The framework developed in this review formalizes this insight and brings reference sensitivity, restoring feedback, and source feedback, which have been treated separately and implicitly in some previous work, under a unified framework. The framework also clearly distinguishes forcing-based sensitivity from effective sensitivity. 

Looking ahead, the field should advance across all three components of the framework. Better estimates and representations of anthropogenic heat forcing remain essential, but equally important are more rigorous characterization of sensitivity and more systematic diagnosis of feedbacks. The following subsections discuss future priorities for advancing each of these three components.

\subsection{Quantifying and representing anthropogenic heat forcing}

For anthropogenic heat forcing, the next challenge is not simply to produce more $Q_F$ estimates, but to make them standardized, accompanied by quantified uncertainties, and physically usable in numerical models. This requires two complementary efforts: standardizing and benchmarking $Q_F$ datasets, and representing $Q_F$ as a physically structured forcing.

\subsubsection{Standardization, benchmarking, and scenario-based future $Q_F$ datasets}

Current global and regional $Q_F$ datasets differ in units, spatiotemporal resolution, sectoral decomposition, and reporting conventions. In addition, deterministic estimations of $Q_F$ should provide adequate uncertainty quantification. Establishing community standards for data formats, metadata, sector definitions, reporting conventions, and uncertainty quantification would therefore facilitate localized evaluations in data-scarce regions, comparison among datasets, and streamlined applications in weather, climate, and earth system models. Similar to the role played by the Emissions Database for Global Atmospheric Research (EDGAR) \cite{edgar_2018} for atmospheric emissions, an operational and regularly updated global anthropogenic heat forcing database would provide a common resource for numerical modeling.

Future research should also place greater emphasis on intercomparison among independent estimation methods. Comparisons among inventories, residual surface energy balance approaches, building energy models, remote sensing products, and eddy covariance observations in well-observed cities provide opportunities to quantify uncertainties in $Q_F$ and identify method-specific biases. The study by \citeA{chow_2014} illustrates the value of such benchmark studies, showing that multiple independent approaches can converge to consistent estimates while also identifying their respective strengths and weaknesses. Expanding similar benchmarking efforts across diverse climates, urban forms, and socioeconomic settings would provide a stronger foundation for evaluating anthropogenic heat flux data products.

\citeA{sailor_methods}'s vision of community-scale anthropogenic heat forcing datasets for atmospheric modeling has largely been realized for historical and present-day conditions, although these products still require further standardization, comprehensive benchmarking, and systematic uncertainty quantification. The next challenge is to move toward scenario-based future $Q_F$ datasets \cite{varquez_2021} that describe how anthropogenic heat flux may evolve under continued urbanization, socioeconomic development, technological change, and climate-driven energy demand. Many sectoral components of $Q_F$, including industrial emissions, transportation, and metabolic heat release, have yet to be distinctly provided for future projection purposes. 

Developing, maintaining, and properly documenting such datasets would allow anthropogenic heat flux estimates to be more consistent with changes in climate, urbanization, and energy systems; and always up-to-date with advancing computational approaches.

\subsubsection{Source and release pathway representation}

Recent advances have enabled explicit representation of building, transportation, industrial, and metabolic heat sources, as well as dynamic coupling between BEMs, traffic modules, and urban land surface models. Future developments should continue this transition from prescribing total anthropogenic heat fluxes toward physically based representations of individual source sectors. 

Of equal importance is the pathway through which anthropogenic heat enters the urban climate system, which may be related to, but is not necessarily identical to, the source sector. Heat released at street level by vehicles, rejected above roofs by HVAC systems, emitted from industrial stacks, or discharged from underground infrastructure can produce different climate responses even with the same total $Q_F$  \cite{Kikegawa2003AE,Li2024Structural,zhang2024relative}. This issue was already recognized by \citeA{oke_1974_wmo} and \citeA{sailor_methods}, yet comparatively little progress has been made in representing these release pathways explicitly. Developing data and parameterizations for these release pathways, including their vertical distribution and physical characteristics, should therefore become a priority for next-generation $Q_F$ datasets and urban climate models.

\subsection{Defining and characterizing anthropogenic heat sensitivity}
\label{subsect:defining-sensitivity}

Compared with anthropogenic heat forcing, anthropogenic heat sensitivity remains less well constrained. Observational estimates are still very limited, while existing model estimates are affected by inconsistent definitions and methodological differences across studies. Although pioneering studies have begun to clarify the typical magnitude, variability, and uncertainty of anthropogenic heat sensitivity, major gaps remain in how sensitivity is defined, computed, and interpreted, and in whether estimates obtained for one model, location, or spatial and temporal scale are applicable to others. Future progress therefore requires stronger observational constraints, greater consistency in how anthropogenic heat sensitivity is defined and reported, and a more systematic understanding of its variability and uncertainty. 

In this review, anthropogenic heat sensitivity is interpreted as a scalar relationship between anthropogenic heat flux and an equilibrium temperature response. In reality, however, both $\Delta Q_F$ and $\Delta T$ vary spatially and temporally, and a temperature change at a given location and time is not necessarily driven solely by the anthropogenic heat flux change at that same location and time. Diagnosing a scalar anthropogenic heat sensitivity therefore depends critically on how the anthropogenic heat flux and response are averaged over specified spatial domains and temporal windows. This raises two important questions. First, how does the diagnosed sensitivity depend on the spatial and temporal scales over which the forcing and response are averaged? Second, does the diagnosed sensitivity depend on the spatial and temporal pattern of anthropogenic heat forcing within the averaging domain and window? These questions are discussed below.

\subsubsection{Characterizing scale dependence}
\label{subsect:scale-dependence}

Scale dependence first arises from the temporal averaging window, as different components of the urban climate system adjust to anthropogenic heat flux perturbation over different timescales. Canopy air temperature may adjust rapidly toward a quasi-equilibrium state, but other processes that influence anthropogenic heat sensitivity may operate on longer timescales. Turbulent exchange between the canopy layer and the overlying atmosphere may act over seconds to minutes, whereas surface temperature and boundary-layer temperature adjustments, and source feedbacks involving building energy use can evolve over much longer periods.

A quasi-equilibrium sensitivity can therefore be meaningful for canopy air temperature when the averaging timescale is long compared with the adjustment timescale of the canopy air subsystem, a condition that is often attainable. However, it should not automatically be interpreted as the equilibrium response of the entire urban surface or atmosphere, which generally requires longer timescales. This distinction is especially important for simulations lasting only several hours or a single day, in which canopy air temperature may approach quasi-equilibrium while slower surface, boundary-layer, and mesoscale adjustments remain incomplete. When the averaging timescale is longer than the canopy-air adjustment timescale but shorter than the adjustment timescales of these slower processes, the diagnosed sensitivity may continue to evolve as those processes respond, manifested as transient or only partially realized feedbacks.

The spatial averaging domain is equally important because anthropogenic heat forcing and temperature response are generally not co-located. Horizontal advection, turbulent transport, and mesoscale circulations allow anthropogenic heat emissions released in one location to influence temperatures elsewhere. Ideally, the averaging domain should encompass the spatial footprint of the temperature response induced by the anthropogenic heat perturbation. In practice, however, this response footprint is rarely quantified and likely depends on factors such as atmospheric stability, wind, boundary-layer depth, urban morphology, and the spatial distribution of anthropogenic heat emissions. Different averaging domains may therefore capture different fractions of the total temperature response, making sensitivities diagnosed at different spatial scales not directly comparable.

Future studies should therefore systematically quantify how diagnosed sensitivities vary with the temporal averaging window, the spatial averaging domain, and the lateral boundary conditions of the model domain, and determine whether they converge beyond characteristic temporal or spatial scales. Such analyses would help establish the relationships among instantaneous, hourly, daily, seasonal, and longer-term sensitivities, as well as among canopy-, neighborhood-, city-, regional-, and global-scale sensitivities. Future studies should also examine how anthropogenic heat sensitivity changes as additional physical processes and feedbacks, including water vapor, clouds, aerosols, and snow and ice, are represented, and extend the forcing–response–feedback framework to other control volumes and state variables (e.g., the surface--canopy--boundary-layer continuum), with explicit consideration of the spatial scale of the forcing perturbation and the adjustment time scale of the system \cite{yang2016urban,DemchenkoGinzburg2018,xue2020impact,khanh2025impact}. %

\subsubsection{Characterizing pattern dependence}

Anthropogenic heat sensitivity may depend not only on the mean anthropogenic heat forcing averaged over the selected temporal window and spatial domain, but also on how that forcing is distributed in time and space.  Temporally, the same mean anthropogenic heat release may occur as a nearly constant background forcing, exhibit strong diurnal variations, or be concentrated during short episodes associated with traffic, building cooling demand, or industrial activity. Spatially, the same domain-mean anthropogenic heat flux may be concentrated within dense urban cores, distributed across residential neighborhoods, or aligned along transportation corridors.

Future studies should determine whether, and under what conditions, temporal and spatial forcing patterns affect anthropogenic heat sensitivity. Rather than relying only on observed patterns of anthropogenic heat flux, idealized experiments could systematically vary the structure of the forcing while holding the total heat input fixed. This would help characterize the response function of the urban climate system and test whether centralized, distributed, corridor-like, constant, diurnally varying, or episodic forcing produces meaningfully different sensitivities. Characterizing the role of forcing pattern would help determine when the mean forcing is sufficient for defining anthropogenic heat sensitivity and when more temporally or spatially resolved descriptions are needed.

If anthropogenic heat sensitivity depends strongly on the spatial or temporal pattern of the forcing, a single scalar sensitivity based on mean $\Delta Q_F$ may be insufficient to characterize the response. More general formulations could instead adopt a source–receptor perspective, representing the temperature response at a given location and time as a function of anthropogenic heat flux across different locations and times. This would allow $\Delta Q_F$ and $\Delta T$ to be evaluated over different spatial domains and temporal windows and could better capture transport, storage, and delayed adjustment. Developing and testing such formulations will require substantial future work, but they may ultimately provide a more general framework for characterizing anthropogenic heat sensitivity across scales.

\subsection{Diagnosing and understanding anthropogenic heat feedbacks}

Despite increasing recognition of the importance of anthropogenic heat feedbacks, their quantitative understanding remains at a much earlier stage than that of anthropogenic heat forcing and sensitivity. Many studies have examined the effects of feedbacks through coupled simulations, often by comparing temperature responses with and without particular feedback pathways, yet relatively few have attempted to quantify the role of these feedbacks using explicit feedback parameters (or equivalent gain factors). Future progress therefore requires moving beyond documenting the temperature and other effects of feedbacks toward diagnosing, isolating, and comparing feedback parameters across different models, cities, climates, and temporal/spatial scales.

\subsubsection{Diagnosing feedback parameters}
\label{subsect:diagnosing-feedback}

A first priority is to define feedback parameters in ways that can be diagnosed consistently from model outputs. Studies such as \citeA{Kikegawa2022AE} and \citeA{Wang2023ERL} represent important steps in this direction because they estimate feedback parameters directly rather than only reporting the temperature effects associated with particular feedback pathways. Nevertheless, diagnosing feedback parameters is not always straightforward. For example, \citeA{Kikegawa2022AE} estimated $\lambda_{\rm source}$ from the regression of $Q_F$ against $T$ within their weekdays-run. However, because temperature covaries with other meteorological controls on building energy demand and thus $Q_F$, this regression should be interpreted as an operational diagnostic rather than a strict estimate of the partial derivative defined in the present framework. Likewise, regression-based estimates from empirical energy demand-temperature relations \cite<e.g.,>{pigeon_2007,ginzburg2021dependence} and building energy simulations \cite<e.g.,>{Takane2019NPJ,Li2024NatClimate} generally do not isolate temperature effects with all other factors held fixed. 

Similarly, $S_0$ is formally defined as the reference sensitivity in the absence of source and restoring feedbacks, whereas many estimates, including that of \citeA{Kikegawa2022AE}, are derived from simulations that retain both feedbacks. Overall, such estimates should be interpreted as operational diagnostics rather than strict representations of the formal quantities defined in the present framework.

A related challenge concerns the definition of forcing-based sensitivity in studies that include source feedback. As emphasized in Sections~\ref{sect:sensitivity} and~\ref{sect:feedback}, once anthropogenic heat flux is allowed to respond to temperature, separating the anthropogenic heat forcing from the temperature-dependent source-feedback contribution is not always straightforward. For example, the total anthropogenic heat flux predicted by a BEM generally contains both a reference heat release that acts as the forcing and an additional temperature-dependent contribution arising from source feedback. One conceptually straightforward way to isolate the source-feedback contribution would be to compare a fully coupled simulation with a counterfactual simulation in which the release of anthropogenic heat into the atmosphere is suppressed. However, suppressing anthropogenic heat release while continuing to simulate building energy consumption effectively introduces an energy sink, rendering the coupled building–atmosphere energy budget physically inconsistent \cite{Kikegawa2022AE}.

This methodological tension reflects different objectives in urban climate modeling and climate feedback analysis, but it also points to a broader opportunity for future research. Methodologies developed in climate science for diagnosing forcing, sensitivity, and feedback provide a valuable conceptual framework for interpreting coupled urban-energy simulations, but they must be adapted to account for the physical constraints of these models. Future work should therefore develop diagnostics and experimental designs that isolate feedback parameters without sacrificing the physical consistency of coupled building–atmosphere interactions.

\subsubsection{Understanding feedback parameters}

Substantially more work is needed to understand how feedback parameters vary across space, time, urban form, and background meteorological conditions. Current knowledge remains limited because only a small number of studies have quantified these parameters explicitly, and some feedbacks have not yet been quantified at all.

A central priority is to quantify the variability of the baseline feedback parameter because it sets the reference against which source and restoring feedbacks are evaluated: the same feedback parameter can imply very different gain factors depending on the magnitude of $\lambda_0$. Future work should systematically characterize the variability of $\lambda_0$ and identify the physical processes that control it.

Existing studies of restoring feedback have focused primarily on processes within the urban canopy \cite{Wang2023ERL}, although \citeA{Li2024Structural} extended the analysis to include atmospheric temperature adjustments in land–atmosphere coupled simulations. Restoring feedbacks associated with broader boundary-layer adjustment, horizontal advection, and mesoscale circulations, however, remain largely unquantified, as do the characteristic timescales over which these feedbacks develop and become fully realized. Extending the framework across the entire urban boundary layer and to larger spatial and temporal scales therefore remains an important next step.

Although the sign of source feedback appears to be relatively well established in many conventional heating and cooling systems, quantitative estimates remain limited, leaving its dependence on climate, building characteristics, energy systems, technology, and human behavior largely unknown. Emerging technologies may also introduce feedback pathways beyond those captured by the conventional definition of anthropogenic heat flux. For example, air-source heat pumps are expected to reduce wintertime $Q_F$ in the traditional sense. However, colder conditions can increase heating demand and associated heat extraction from the outdoor air, thereby forming an amplifying local winter-cooling pathway in building–atmosphere heat exchange  \cite{Meyer2024NatCommun,Takane2024GMD,Xie2024SCS}. This pathway is conceptually similar to a positive source feedback, although it has not generally been quantified explicitly using a feedback parameter. Quantifying both conventional source feedbacks and these new feedback pathways will be essential for projecting future urban climate responses as cities electrify, building technologies improve, cooling demand increases, and patterns of energy consumption change.

\subsubsection{Beyond linear feedback theory}
\label{subsect:beyond-linear-feedback-theory}

Finally, the forcing–response–feedback framework developed in this review is based on a local linearization of the urban climate system, whose range of validity requires much more systematic investigation. Existing studies provide mixed evidence. Offline land-model simulations have reported declining sensitivities under large anthropogenic heat flux perturbations \cite{Wang2023ERL}, whereas regional climate simulations have found similar sensitivities over a range of $Q_F$ values, implying approximately linear temperature responses \cite{Li2024Structural}. More recently, idealized large-eddy simulations under calm conditions demonstrated power-law rather than linear scaling between urban temperature and surface heat flux, indicating nonlinearities arising from the atmospheric response \cite{o2026unifying}. 
Similarly, global modeling studies have suggested that the strength of source feedback may itself change under a warming climate \cite{Li2024NatClimate}. 
Together, these results suggest that the validity of local linearization depends on both the magnitude of the forcing and the state of the urban climate system, and may also depend on the spatial and temporal scales over which the response is considered. Future work should identify the conditions under which linear theory remains applicable, determine when nonlinear behavior becomes important, and develop generalized forcing–response frameworks capable of describing both linear and nonlinear regimes.

\section{Final Remarks}

Anthropogenic heat flux ($Q_F$) has long been recognized by the urban climate community as an important component of the urban surface energy balance, with potential relevance to the broader climate science community. However, relatively little work has sought to interpret the role of anthropogenic heat flux using the conceptual language of forcing, sensitivity, and feedback that underpins much of modern climate science, as the urban climate community has largely approached anthropogenic heat flux from an applied perspective. This review represents a step toward bridging the gap between applied urban meteorology and climate science by introducing a forcing–response–feedback framework for understanding the temperature effects of anthropogenic heat flux.

A central message of this review is that the climatic significance of anthropogenic heat flux depends not only on the forcing itself, but also on the response of urban temperatures to that forcing and on the feedbacks that amplify or damp the response. Reinterpreting the existing literature through this framework reveals that many seemingly disparate studies share common physical interpretations, while also highlighting ambiguities in the definition and diagnosis of anthropogenic heat sensitivity and feedback parameters that complicate comparisons across studies.

This review synthesizes selected studies that quantify anthropogenic heat sensitivities and feedback parameters, and uses them to distinguish forcing-based sensitivity, effective sensitivity, and reference sensitivity. It also clarifies the differences among baseline, restoring, and source feedbacks. More broadly, it demonstrates how methodologies developed in climate science can provide valuable guidance for interpreting urban climate studies, while also emphasizing the need to adapt these methodologies to accommodate the physical constraints of urban climate models.

Finally, this review suggests that future progress depends not only on improving estimates of anthropogenic heat forcing, but also on systematically quantifying sensitivities and feedbacks across different cities, climates, and modeling frameworks. Developing transferable physical understanding requires giving as much attention to the forcing–response relationship as to the forcing itself. We hope this review demonstrates the value of bringing together concepts from climate feedback theory with urban observations and simulations, and motivates similar efforts in other domains, including urban climate adaptation, where the effects of interventions such as reflective roofs may likewise be interpreted through the lens of forcing, sensitivity, and feedback \cite{li2025bridging}.

\begin{glossary}

\item[Anthropogenic heat flux ($Q_F$)] Heat released to the environment by human activities, including buildings, transportation, industry, and metabolism, expressed as an energy flux per unit area. Related terms used in this review include anthropogenic heat emissions, heat release, anthropogenic heating, and waste heat.

\item[Anthropogenic heat forcing] Anthropogenic heat flux considered as an energy input that perturbs the urban climate system and initiates a climatic response.

\item[Anthropogenic heat sensitivity] The temperature response per unit anthropogenic heat flux.

\item[Reference sensitivity ($S_0$)] Anthropogenic heat sensitivity in the absence of additional restoring and source feedbacks, with only the baseline response retained. 

\item[Forcing-based sensitivity ($S_f$)] Anthropogenic heat sensitivity defined using the anthropogenic heat forcing as the denominator.

\item[Effective sensitivity ($S_e$)] Anthropogenic heat sensitivity defined using the total anthropogenic heat flux change, including contributions from source feedback, as the denominator.

\item[Feedback] A process through which an initial temperature response alters energy fluxes, thereby modifying the temperature response.

\item[Baseline feedback or baseline response] The restoring tendency of the urban climate system in the reference state, before additional restoring or source feedbacks are considered.

\item[Restoring feedback] Feedback arising from temperature-dependent changes in radiative, turbulent, or other energy exchanges.

\item[Source feedback] Feedback in which anthropogenic heat flux changes in response to temperature, such as increased air-conditioning demand.

\item[Feedback parameter] The change in energy flux associated with a feedback process per unit temperature change, representing the strength and direction of that feedback.

\item[Baseline feedback parameter] The change in energy flux per unit temperature change associated with the baseline response. The inverse of its absolute value is the reference sensitivity.

\item[Gain factor] A dimensionless measure of temperature amplification or damping due to a feedback, defined as the product of that feedback parameter and the reference sensitivity.

\item[Building energy model (BEM)] A model that simulates building energy use and associated anthropogenic heat emissions.

\item[Urban canopy model (UCM)] A model representing exchanges of momentum, heat, and radiation among urban surfaces and the urban atmosphere.

\item[Urban heat island (UHI)] The tendency for urban areas to be warmer than their rural surroundings, often quantified by the urban–rural temperature difference.

\end{glossary}

\section*{Open Research Section}

The materials needed to reproduce the corpus assembly and structured extraction are available in \citeA{li2026_rog_ah_archive} and explained in the supporting information. The corpus and its structured extraction can also be browsed through a companion website (\url{https://ah.urbanclimaterisk.group}), which supplements the archived materials in \citeA{li2026_rog_ah_archive} but does not replace them as the citable record of the data supporting this review. Data and scripts for reproducing the figures are also available in \citeA{li2026_rog_ah_archive}. The global datasets of $Q_{F}$ presented in Figure \ref{fig:global_comparisons} are accessible via publicly-available repositories and archives. Specifically, AH4GUC \cite{varquez_2021}, AH-DMSP \cite{ah_dmsp_2017}, PF-AHF \cite{pf_ahf_2019}, and Dong \cite{dong_2017} are available at \url{https://doi.org/10.6084/m9.figshare.12612458}, \url{https://doi.org/10.6084/m9.figshare.c.3647903}, \url{https://doi.org/10.6084/m9.figshare.c.4182824}, and \url{https://doi.org/10.5281/zenodo.21767353}, respectively. Figure \ref{fig:global_comparisons} uses the 2013 field of the $0.5^{\circ}$ dataset of \citeA{flanner_2009}, which provides annual fields for 2005--2040 (\texttt{AHF\_2005-2040\_0.5x0.5.nc}). That dataset is no longer distributed by its original host, \url{http://www.cgd.ucar.edu/tss/ahf/data/}, which went offline in 2024, and it has no persistent identifier of its own; the copy used here was retrieved on 7 August 2026 from the Internet Archive's capture of 13 April 2012, verified against the checksum that the archive records for that capture, and is redistributed together with its provenance record in \citeA{li2026_rog_ah_archive}.

\section*{Conflict of Interest Statement}
The authors declare there are no conflicts of interest for this manuscript.

\acknowledgments
We thank Dr. Yukihiro Kikegawa and Dr. Do Ngoc Khanh for their insightful discussions and valuable perspectives, and three anonymous reviewers for their careful reading and constructive comments, which substantially improved this review. We also thank Xuanzong Zhang for preparing the theory figure and Dr. Linying Wang for generously providing the data used in the feedback synthesis figure. This review was conceived and largely written during DL’s visit to the Institute of Science Tokyo, which was supported by Boston University and the World Research Hub Program of the Institute of Science Tokyo, and benefited greatly from the stimulating discussions and collaborative environment there. DL also acknowledges partial support from the U.S. Department of Energy, Office of Science, as part of research in MultiSector Dynamics, Earth and Environmental System Modeling Program. ACGV acknowledges the support of JSPS Kakenhi grant JP25K00040. TS acknowledges support from the Royal Society--NSFC International Exchanges scheme (Grant No. IEC\textbackslash NSFC\textbackslash 242040). YT acknowledges support from the Japan Science and Technology Agency (JST) FOREST Program (Grant Number JPMJFR241G, Japan) and the Environment Research and Technology Development Fund (JPMEERF25S12433) of the Environmental Restoration and Conservation Agency provided by the Ministry of the Environment of Japan. JY acknowledges support from the National Natural Science Foundation of China (Excellent Young Scientists Fund, Grant No. 42322903). DJS acknowledges partial support from the U.S. Department of Energy, Office of Science, Office of Biological and Environmental Research’s Urban Integrated Field Laboratories research activity, under Award Number DE-SC0023520.

\bibliography{agusample_clean}

\clearpage


\section*{Supplementary material}

## Contents of this file

## Introduction

This review builds upon a compiled corpus of 577 papers. The primary purpose of this Supporting Information (SI) is to document how this corpus was compiled (Text S1; Figure S1; Table S1) and to summarize the insights gained (Text S2). The SI is intended to provide complete transparency so that the corpus can be retraced, independently examined, and, where appropriate, reproduced.

The corpus was assembled with the help of a large language model (LLM). The primary motivation was scale: an LLM can assist in screening the thousands of records returned by Web of Science searches, identifying those relevant to the scope of this review, and then extracting structured information from the retained papers. Throughout the project we used a single LLM family, Claude Opus by Anthropic, adopting the most capable versions available during the project period (4.7 and 4.8). The newer version was adopted as it became available part-way through the project, and we did not observe systematic differences between the two versions for these tasks. Using the most capable models was important for enabling the workflow, but the rigor of the results was ensured by the surrounding procedures, including screening audits, human adjudication, and the sign-off process discussed later. Through this process, we demonstrate how modern artificial intelligence (AI) tools can assist a small research team in screening, organizing, and synthesizing a large body of literature, while also documenting the lessons learned, limitations encountered, and best practices that emerged from this exercise. In this SI, *reviewer* denotes an author performing an evaluation or adjudication task within the LLM-assisted workflow described in Text S1, rather than a journal peer reviewer.

Beyond documenting the corpus and its compilation, this SI also provides supporting information for the study-level anthropogenic heat sensitivity synthesis presented in the main text. Table S2 lists the studies included in that figure and maps each abbreviated study label to the corresponding study title and DOI. It also reports the number of sensitivity estimates extracted from each study, the study-level geometric mean sensitivity, and the within-study range shown in the figure.

## Text S1. LLM-assisted corpus assembly and information extraction

### S1.1. Web of Science search strategy

We started by searching the Web of Science Core Collection in two stages: first, a exploratory query was conducted on 21 April 2026, followed by the finalized query on 22 April 2026. The date defines the reproducible search snapshot, because database coverage, indexing, and document metadata can change after a query is completed.

The exploratory query involved multiple steps. It started with a baseline eight-branch query (Table S1), covering modern terminology related to anthropogenic heat and waste heat in atmospheric, weather, and climate contexts. Then, it was expanded with 12 additional branches, representing historical and adjacent terminology identified primarily from pre-2005 source papers. The resulting 20-branch query returned 5,243 records, of which 2,425 were not retrieved by the baseline eight-branch query.

A manual title-and-abstract review of these 2,425 records found that approximately 80% were irrelevant to the scope of this review, including studies of aquatic thermal discharge, engines, and heat-recovery engineering. We therefore decided to retain only 8 of these 12 branches in the finalized query. Of these 8 branches, 6 were kept without modification

because they yielded a meaningful outcome, whereas the remaining 2 were narrowed by adding urban or atmospheric context terms.

Thus, the finalized query comprised 16 branches: the 8 baseline branches and 8 additional branches (see Table S1, which also reports the number of records returned by each branch), returning 3,038 records in Web of Science. After de-duplication and minor record cleanup, 3,035 records were retained as input for the next step.

At this stage, we favored breadth over precision, accepting many irrelevant records to reduce the risk of missing relevant ones. We also did not apply document-type filter, so conference proceedings and other non-article records were retained.

**S1.2. LLM title–abstract screening**

We then screened each record with the help of an LLM according to two main rules. First, a record was retained only if it presented substantive evidence on at least one of the three topics in the forcing–response–feedback framework: *forcing* (topic I), *sensitivity* (topic II), or *feedback* (topic III). Records spanning multiple topics were assigned the corresponding combinations (I+II, II+III, I+III, or *all*), whereas records mentioning anthropogenic heat flux only incidentally were labeled *others*, and out-of-scope records were labeled *invalid*. Second, topic labels were assigned solely on the basis of evidence reported in the paper itself, rather than its framing or citations.

The 3,035 records from the finalized query were screened on 24 April 2026 using Claude Opus (version 4.7), based on the title and abstract. For each record, the LLM returned an inclusion decision (i.e., whether to include or exclude this record), a topic label, a confidence level (high, medium, or low), and a brief explanation of the rationale. The confidence level reflects the LLM's assessment of how clearly the title and abstract satisfy

the rules discussed above: *high* when the evidence is explicit, *medium* when it is suggestive but inconclusive, and *low* when the available information is limited.

The screening process and associated prompts were refined iteratively through testing and comparison with author judgments, particularly to distinguish papers that estimated anthropogenic heat flux from those that merely cited existing estimates and to clarify the treatment of studies that did not examine air temperature effects. The complete prompt history is archived (see Text S1.9) (Li et al., 2026).

Among the 3,035 records, 513 records were recommended by the LLM for inclusion and 2,522 for rejection. Among the 513 recommendations for inclusion, 276 were assigned high confidence, 229 medium confidence, and 8 low confidence; among the 2,522 recommendations for rejection, 2,497 were assigned high or medium confidence, and 25 low confidence.

To compare the LLM results with author judgment, two authors independently labeled a 93-record reference set (= 44 + 49). The LLM results agreed with the reference set for 74 of the 93 records (79.6%), with similar agreement across the two subsets (35/44 and 39/49). Because the reference set also informed the prompt refinement, this agreement should be interpreted as an internal consistency check rather than an independent estimate of the LLM's screening accuracy.

### S1.3. Human adjudication and reconciliation

The 276 records recommended for inclusion with high confidence were retained without human adjudication at this stage. The reliability of high-confidence screening calls is supported by the recall audit described later in Text S1.5. Human adjudication therefore focused on the 262 less certain cases: 237 (= 229 + 8) records recommended for inclu-

sion with medium or low confidence and 25 records recommended for rejection with low confidence.

All 262 records were independently reviewed by two reviewers using a web-based interface that displayed the title, abstract, and LLM results, together with an accept/reject decision form and, where relevant, an option to correct the topic label assigned by the LLM. This stage was deliberately not blinded: reviewers could see the LLM output because the objective was not to perform a fully independent human review of all records, but rather to focus expert attention on the relatively small set of records for which the LLM was less confident.

Among the 262 records, the two reviewers reached agreement on 180 records, defined as the same inclusion decision and topic label; 156 of these were retained. The remaining 82 records were reconciled by considering the LLM's decision, both reviewers' notes, and an independent reassessment by a third reviewer. In the end, 64 of these 82 records were retained. For each record, the final decision and its basis were documented. The reconciliation was completed and locked on 4 May 2026. Combining these 220 (= 156+64) manually reviewed records with the 276 high-confidence records yielded a collection of 496 papers, which is referred to as the *reconciled shortlist*.

Note that the topic assigned to each record at this stage is referred to as the *reconciled topic label*. The extraction step described later (Text S1.6) may identify evidence for different topics within individual parts of a paper, so its results may not always match the *reconciled topic label*.

## S1.4. Citation backfill through backward citation chasing

Older studies often used terminology that differs from the modern language of "anthropogenic heat" and are not consistently indexed in current abstract databases. As a result, even a carefully designed keyword search is likely to miss part of this literature. We therefore supplemented the database search by examining the reference lists of papers in the reconciled shortlist and tracing citations back to earlier studies, a process referred to as *citation backfill*, with particular attention to pre-2000 work.

The citation backfill was conducted manually. For each candidate reference, the authors reviewed the source, extracted its bibliographic information, standardized the metadata, checked for duplicates in the corpus, screened with the same rules, and retained it after applying the same adjudication and reconciliation procedure discussed in Text S1.3. This process added 48 candidate references, of which 1 was later removed as a duplicate, resulting in a *backfill-expanded shortlist* of 543 papers.

**S1.5. Re-screening and recall audit of rejected records**

During human adjudication and reconciliation (Text S1.3), we reviewed all records recommended for inclusion with medium or low confidence, as well as all low-confidence rejections. The medium- and high-confidence rejections were therefore the only records that had not yet received human review. We re-screened and audited these records as described below.

We first re-screened the 2,497 medium- and high-confidence rejected records with the same LLM procedure and rules described in Text S1.2. This re-screening procedure identified 154 records for further review and rejected 2,343 records. Applying the inclusion criteria reduced the 154 records to 28 candidates, of which 25 were retained after full-text assessment.

We next used a stratified random audit to estimate how many relevant papers might still have been missed. The rejected records were grouped by the model's reported confidence, with a larger sampling fraction assigned to the more uncertain medium-confidence group. Using a fixed random seed for reproducibility, we sampled 150 of the 2,343 records rejected during re-screening: 90 high-confidence and 60 medium-confidence rejections.

None of the 90 high-confidence rejections was found to be relevant. This corresponds to a Wilson 95% upper confidence bound of approximately 4% for the miss rate in this group. We therefore accepted the high-confidence rejections without further review.

Among the 60 medium-confidence rejections sampled, 9 were flagged for further review. Although none of the 9 records was confirmed as a relevant paper after full-text assessment, this number was large enough to justify reviewing all remaining medium-confidence rejections. Re-screening based on their titles and abstracts identified 26 further candidates; 12 were retained after full-text review.

After this step, the provisional corpus comprised 580 papers: 543 in the backfill-expanded shortlist, 25 recovered through re-screening, and 12 added through the recall audit.

### S1.6. Structured information extraction

The forcing–response–feedback framework developed in this review article relies on specific information such as reported $Q_F$ values, temperature responses, and feedback mechanisms. Manual collection of these specifics would be slow, uneven, and hard to audit. This is where LLM assistance becomes particularly useful. An LLM can read every full text against a predefined, version-controlled schema (that is, a defined list of fields to extract, each with a specified format) and return structured, machine-checkable informa-

tion. This helps keep the extraction consistent, transparent, and easy to audit. Moreover, because each paper was associated with a PDF, the LLM could access the full article, including figures, tables, and captions, in addition to the main text. This was particularly important because relevant evidence was often distributed across different components of a paper, including the main text, figures, tables, and captions.

The extracted results included separate sections for *forcing*, *sensitivity*, and *feedback*. The forcing section captured reported $Q_F$ values, units, spatial and temporal scales, estimation methods, and sectors. The sensitivity section recorded the temperature variable, magnitude, method, and spatial scale. The feedback section described the feedback mechanism, sign, whether $Q_F$ was prescribed or prognostic, and whether the paper diagnosed a restoring feedback. Each section was assigned a status of *found*, *partial*, or *not in paper*. Only sections assigned *found* were treated as providing strong evidence. A *partial* status indicated that the paper discussed or anticipated the topic without fully establishing it and was therefore treated as weak evidence. Note again that these extracted results do not replace the *reconciled topic label* discussed earlier (Text S1.3). The two may differ because they are based on information from different sources (one from title/abstract and one from full text).

The extraction was not completed in a single step. It began with a pilot extraction on 13 May 2026, followed by the first systematic extraction from 14 May 2026 to 15 May 2026. The same procedure was subsequently extended to late additions as they entered the corpus, including papers recovered through re-screening and recall audits (Text S1.5) and four additional papers recovered during later review (discussed below). Every extraction run, together with its prompt version and model, is archived (Text S1.9).

The extraction was subsequently refined together with continued reviewer sign-off (to be discussed in Text S1.7), with different sections refined differently. The forcing section was unchanged, with newly identified information recorded in notes. The feedback section underwent repeated refinement as our understanding of what constituted valid feedback evidence evolved. The first refinement (referred to as loop-taxonomy reread) separated source and restoring feedbacks, the second refinement (referred to as quantified-parameter criterion) required a feedback parameter, and the last refinement (referred to as admissibility gate) classified feedback as *found* only when a paper directly reported a feedback parameter or provided sufficient information to derive one from the reported quantities. Feedback parameters clearing the admissibility gate were subsequently normalized to canonical units (Text S1.8). The sensitivity section also underwent three refinements with substantial reviewer involvement (Text S1.7): first, to extract individual numerical estimates together with other relevant information; second, to normalize the sensitivity values to canonical units, namely, K $(\mathrm{W\ m^{-2}})^{-1}$, while preserving their original values and units; and third, to check reported sensitivities against independent quantities reported in the same paper. These refinements supported the subsequent verification process (Text S1.8) while preserving the original extraction records.

Before concluding this subsection, two additional points require clarification. First, a naming difference remains in the archived materials. The first extracted section is referred to as *forcing* in this SI and the main text, but it retains the original name *magnitude* in the extraction schema, archived model outputs, and sign-off records. This was because the schema was created before the review article settled on *forcing* as the section name. The archived files were not renamed retrospectively in order to preserve the original audit

trail. Consequently, variables beginning with *magnitude* in the released data correspond to the *forcing* section.

Second, we explain how the final paper count for the structured extraction was obtained. The preceding steps produced a provisional set of 580 papers. Because the structured extraction relied on full text, we removed 3 records for which no usable full text was available, including one with a PDF that contained no extractable evidence. We also removed 4 citation-backfill records for which neither an abstract nor a full text could be obtained through institutional access or interlibrary loan. 4 papers were then added and extracted using the same procedure: 2 papers that entered the corpus once their full texts were finally retrieved through institutional subscription, and 2 papers published while this review was being prepared, which were added manually. These changes yielded the locked evidence corpus of 577 papers for structured information extraction. 3 systematic reviews of anthropogenic heat (Sailor, 2011; Lu et al., 2024; Feng et al., 2025) were tracked separately as a companion set and were not included in the corpus.

## S1.7. Reviewer sign-off of the extraction

The structured extraction was reviewed by the authors through a web interface that displayed each paper's full text alongside its extracted information. Papers were assigned to reviewers according to their expertise. Reviewers could endorse, modify, or reject the extracted evidence and add a note. All review decisions were documented to provide a complete audit trail. When a paper was reviewed more than once, the most recent decision replaced the earlier one. In total, the reviewer sign-off recorded 959 section-level decisions, which exceeds the number of papers because each paper received separate decisions for each of its sections.

The reviewer sign-off was not only a quality-control step but also a source of methodological refinement. As reviewers examined extracted information, their understanding of what constituted valid evidence evolved, revealing limitations in the original extraction framework. More broadly, as evidence accumulated and our conceptual framework evolved, the structure and organization of the review article were adjusted, requiring refinements to the extraction.

These refinements were implemented differently across different sections of the extraction. The forcing section primarily involved collecting reported evidence, and newly identified information could be incorporated in the notes. Therefore, it did not require repeated re-extraction. In contrast, the sensitivity and feedback sections involved quantitative synthesis and therefore required more iterative refinement. For feedback, this refinement was implemented mainly through progressively stricter definitions of what constituted admissible evidence, leading to multiple re-extractions. For sensitivity, the challenge was not only defining admissible evidence but also obtaining reliable numerical estimates, which required substantial human involvement in interpreting texts, digitizing reported values, and applying the evolving criteria. Throughout this process, insights from reviewer sign-off decisions were incorporated into subsequent extraction steps, allowing the extraction criteria to evolve alongside the scientific framework of the review.

### S1.8. Verification and normalization

As discussed above, the sensitivity and feedback sections required more iterative refinement than the forcing section because they involved quantitative synthesis and required substantial human interpretation. Therefore, these two sections underwent additional verification.

For sensitivity values, reviewer sign-off resulted in 129 candidates identified through both LLM extraction and human review of the full text. These candidate values then underwent three rounds of verification: LLM-based verification, review by an independent reviewer, and a final reviewer check. After verification, the final study-level synthesis included 81 individual estimates from 44 papers (Table S2), all normalized to a canonical unit of K (W $m^{-2}$)$^{-1}$.

The feedback parameters underwent a similar verification and normalization process but at a smaller scale. The final 13 papers that cleared the admissibility gate were normalized to canonical units (W $m^{-2}$ $K^{-1}$ for feedback parameters and dimensionless for gain factors; see the main text), and the feedback figure in the main text is built from these records.

**S1.9. Reproducibility and data availability**

The project was maintained in a version-controlled GitHub repository, while reviewer decisions were collected through a web interface and recorded in the repository. The workflow was designed to be auditable at every step. The counts reported in this SI are not maintained by hand: they are derived from the archived records by an archived script and enforced by automated consistency checks. The accompanying dataset publication (Li et al., 2026) archives the materials needed to reproduce and examine the workflow.

LLM runs were archived with their inputs, outputs, model identifiers, and prompt versions, and each prompt revision was preserved as a separate file. Because the LLM was accessed through a hosted service, repeating the same request may not produce an identical response. Re-running the archived prompts and inputs should reproduce the substance of the results, but not necessarily the exact wording. The archived outputs therefore provide the definitive record of each run.

## Text S2. Lessons from the AI-assisted workflow

In our review, we used AI to support a rigorous, multi-author literature review from initial screening through information extraction and final synthesis, rather than merely assisting an individual researcher with coding or exploratory analyses. To our knowledge, reviews assembled in this manner remain rare. Therefore, we conclude this SI by highlighting three key lessons learned and the remaining unresolved issues, with the aim of enabling other teams to evaluate, adapt, and critically assess this workflow.

### S2.1. Version control is the essential guardrail

Integrating AI into a scientific workflow introduces a fundamental challenge: AI systems are fast and scalable, but their mistakes can be difficult to detect and downstream analyses might have been built upon them by the time these mistakes are discovered. Similar issues also arise from human errors, although the speed and scale of propagation may differ. Version control provides a safeguard against both: not because it prevents mistakes, but because it limits their impact by preserving a traceable history of changes, as it has long done in software development where Git is now standard practice.

As an example, an early version of the extraction prompt classified papers as containing evidence of anthropogenic heat feedback too broadly, and a manual review of a small subset of papers exposed the problem. The correction was implemented as a new prompt version rather than a modification: affected papers were re-extracted under the revised prompt, each result was linked to the prompt version that generated it, and the superseded run was retained alongside the updated one. The re-analysis even identified three papers requiring reclassification. The original error, the correction, and the evidence supporting the correction were all preserved in the record.

Interestingly, the most consequential silent error in the project originated from a human process rather than from the AI system. A summary statistic derived from the corpus was miscalculated and remained unnoticed for months. However, because every intermediate state had been preserved, the sequence of changes that caused the error could be traced and corrected. We subsequently consolidated all summary statistics into a single authoritative file validated by automatic checks, ensuring that future inconsistencies can be detected as part of the recorded workflow.

These examples highlight that a version-control system is most valuable when it is continuous, capturing not only final outputs but also the intermediate states that lead to them. Scientific publications typically preserve successful analyses while leaving behind discarded approaches, revised decisions, and failed attempts. This selective documentation is partly a consequence of the substantial effort required to maintain complete records, causing important context to disappear simply because it was never recorded. Working with AI helps reduce this documentation burden because effective AI-assisted workflows require explicit prompts, schemas, and model specifications, which are naturally preserved alongside each run.

A version-control system is also most valuable when it is accessible to all contributors. Git provides a powerful framework for tracking changes, but its steep learning curve makes it an impractical interface for many researchers, particularly those without prior experience using it. In a collaborative review involving contributors with different technical backgrounds, requiring everyone to interact directly through Git would create a barrier to participation. Conversely, allowing contributors to work outside the shared record would recreate the very problem that version control is intended to prevent: silent and

unrecoverable changes outside the workflow. The challenge is therefore not only to maintain a version-controlled record but also to make that record usable by all participants. To address this challenge, we built a companion web interface that provides access to the same version-controlled record, allowing reviewers to read a paper alongside its extraction, submit decisions, and automatically incorporate those decisions into the shared history.

In summary, this project was designed around version control from the start: one GitHub repository holds everything, including the corpus, search queries, prompts, schema, every AI run, and every review decision. The evidence reported in this manuscript can be traced to the paper from which it originated, the run that produced it, and the reviewer who validated it. The same record keeps human and machine contributions synchronized: the manuscript, schema, and AI prompts evolve together, ensuring that all components operate from the same shared state.

### S2.2. Delegate what we can judge well

The record-keeping function described in the previous lesson is an important but largely infrastructural use of AI. The more consequential step is allowing AI to participate directly in research tasks, such as reading papers, extracting information, and evaluating evidence. Such delegation is appropriate only when two conditions are met: (1) the AI is capable of performing the task, and (2) the researchers understand the task well enough to judge both the output and the way it is produced. The second condition carries the real weight: effective oversight is more than inspecting outputs, because a researcher must know in advance what a scientifically valid and useful result looks like, and what a sound path to that result involves.

As an example, screening was a core task in this project, and it satisfied both conditions for AI delegation. Each paper was judged based only on its title and abstract, a task well within the AI's capability. Indeed, applying a consistent inclusion criterion across thousands of records is precisely where maintaining consistent human judgment becomes challenging, whereas an AI system can apply the same rule consistently at scale. Each AI decision could be evaluated by a reviewer because the inclusion criteria were sufficiently clear and accessible. Although every decision could have been manually verified, doing so would largely negate the efficiency gains of AI-assisted screening. We therefore used targeted auditing to evaluate screening performance. Among 90 sampled high-confidence rejections, none were found to meet the inclusion criteria, placing the residual miss rate below 4% (Text S1.5). This example illustrates that AI delegation is appropriate when researchers possess sufficient understanding of the task to evaluate AI outputs.

In contrast, the task of identifying and evaluating anthropogenic heat feedback evidence initially failed the second condition. At the beginning of the extraction process, we asked the AI to identify papers that studied anthropogenic heat feedback and provide a qualitative description of the reported feedback mechanisms. Although the AI could successfully identify many papers discussing feedback-like effects, we lacked sufficiently clear criteria for determining whether these papers met the scope of this review. After developing the forcing–response–feedback framework, we narrowed the focus to papers that directly reported a feedback parameter or provided sufficient information to derive one from the reported quantities, which established a clear criterion against which AI outputs could be evaluated. This example illustrates that tasks should not be delegated to AI until humans have a clear basis for evaluating the output. Just as a researcher who

cannot assess code quality should not blindly rely on AI-generated code, scientific tasks should only be delegated when researchers understand what constitutes a scientifically valid and useful result within the scope of their research.

### S2.3. Human in the loop should improve the loop

Once a task is delegated to AI, humans should remain in the loop, not only to provide oversight and assess the quality of outputs, but more importantly, to use their observations to improve the workflow itself. If a person is involved only to inspect each AI output, the cost grows linearly with the corpus size and the workflow gains little from the interaction. If, instead, a person's observation changes how the AI operates in subsequent iterations, each review decision becomes an opportunity to improve the workflow.

One example was the extraction of sensitivity values from the papers. Early in the extraction process, a reviewer auditing the sensitivity results identified papers that had been assigned sensitivity values despite ambiguous evidence. In some cases, the paper described a temperature response associated with urbanization broadly or with a combination of urban parameters, without isolating the contribution of anthropogenic heat flux. This observation led to the development of criteria defining whether a reported sensitivity value could be considered attributable to anthropogenic heat flux. These criteria included requirements on the reported temperature response, the attribution to anthropogenic heat flux rather than broader urban effects, and consistency between the spatial and temporal scales of the temperature response and anthropogenic heat flux perturbation. The criteria were incorporated directly into the extraction prompt, allowing subsequent extraction runs to apply them systematically. The result was not merely a correction to individual records, but an improvement to the instrument itself.

The same mechanism operated in other parts of the project as well. For example, human review revealed that the original schema did not distinguish whether anthropogenic heat flux was estimated within the paper or simply adopted from cited references. This distinction was subsequently added as a required field for all extractions. The cumulative impact of this iterative human review was substantial in our project, with 627 of the 959 reviewer sign-off decisions (65%) resulting in changes to the extraction. The ability to propagate such improvements across the corpus is a major advantage of AI-assisted workflows. Once incorporated into a prompt or schema, a revised criterion can be applied consistently across thousands of papers, reducing the possibility of inconsistent interpretation as rules evolve. At the same time, systematic AI errors can reveal where the underlying criteria remain insufficient: repeated error patterns indicate deficiencies in the instrument itself rather than isolated mistakes in individual records, allowing targeted improvements to the workflow.

### S2.4. Closing remarks

We recognize that this human-in-the-loop workflow has limitations: AI systems remain partially opaque, runs may not be perfectly reproducible, and human oversight remains essential. The transferability of this approach also remains uncertain because the experience reported here represents a single review team and a single research topic. Nevertheless, the underlying principle appears broadly applicable: route papers by confidence, involve reviewers where scientific judgment is required, record the reasoning behind decisions, and allow those decisions to improve the instrument as well as the outputs. In this model, AI does not replace expert judgment, but extends the reach of expert judgment by applying continuously refined criteria across large bodies of literature.

We are cautiously optimistic about this way of working, not only because of the work it enabled, but also because it suggests an emerging research paradigm based on increasingly capable human–AI collaboration. This workflow enabled a small research team to accomplish work at a scale that would previously have been difficult to achieve. The main investment is not the tooling itself, which can often be adopted within days, but the process of learning how to work effectively with AI as its capabilities continue to evolve. The knowledge and experience gained through this process will continue to carry forward.

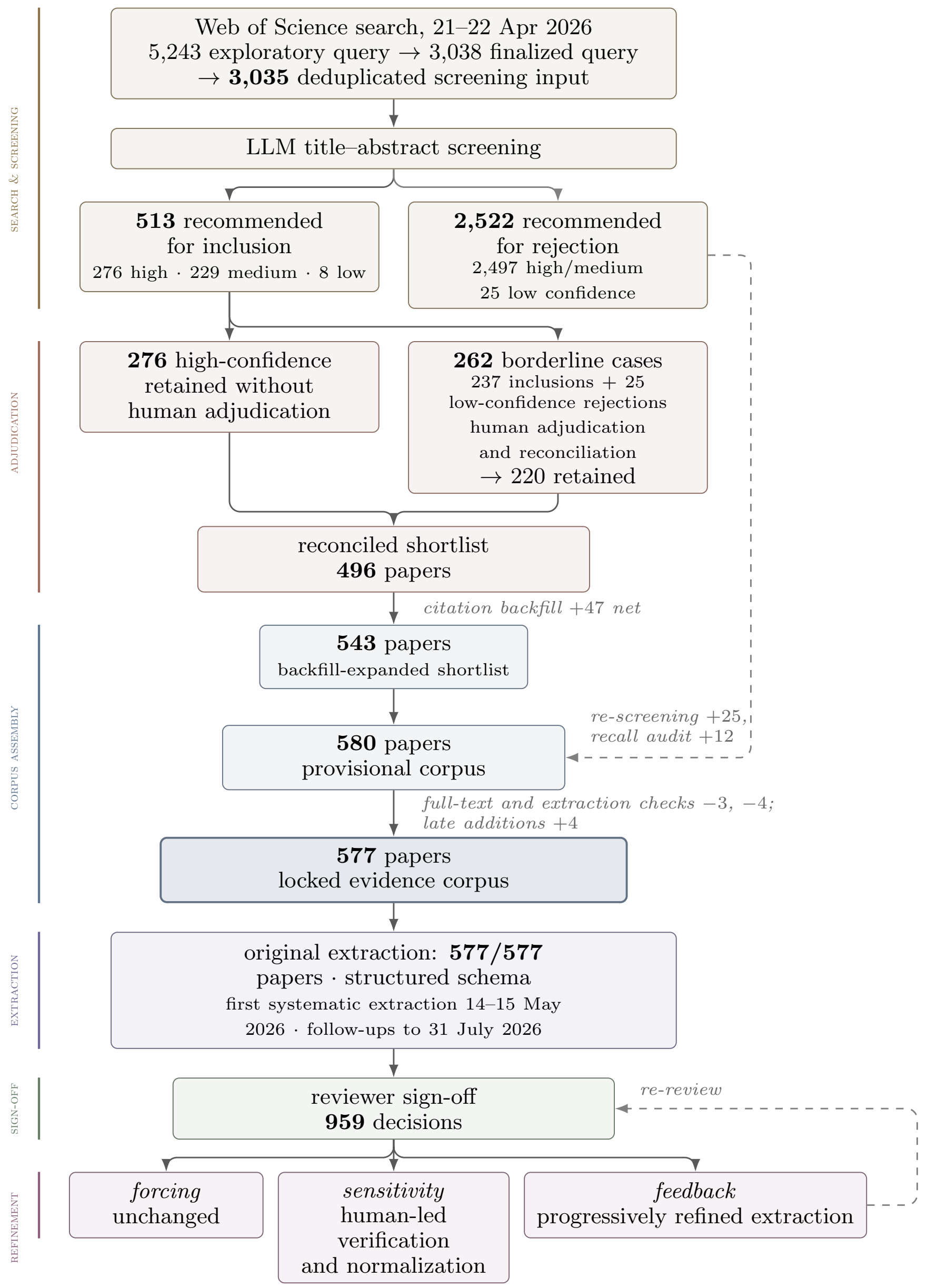


**Figure S1.** Record flow through the LLM-assisted workflow. The 577-paper locked evidence corpus undergoes original extraction, reviewer sign-off, and progressively refined extraction. The dashed re-review path returns refined records to reviewer sign-off; section-specific checks are described in Text S1.8 and run metadata in Text S1.9.

**TABLE S1. The final 16 branches of the Web of Science query (22 April 2026). For each additional branch, *beyond baseline* is the number of records not already returned by the 8 baseline branches. Branches overlap, so the standalone counts sum to more than the combined-query totals shown in the summary rows.**

| Branch | Records | Beyond baseline |
|---|---|---|
| *Baseline branches* | | |
| "anthropogenic heat" | 1,206 | — |
| "anthropogenic heating" | 64 | — |
| "waste heat" AND "atmosphere" | 399 | — |
| "waste energy" AND "atmosphere" | 31 | — |
| "waste heat" AND "weather" | 247 | — |
| "waste energy" AND "weather" | 21 | — |
| "waste heat" AND "climate" | 931 | — |
| "waste energy" AND "climate" | 98 | — |
| *Baseline branches combined, de-duplicated* | 2,818 | |
| *Additional branches* | | |
| "artificial heat" AND (urban OR city OR atmosphere OR climate OR heat island) | 56 | 54 |
| "man-made heat" OR "manmade heat" | 6 | 5 |
| "heat discharge" AND (urban OR city OR building OR anthropogenic OR atmosphere) | 52 | 27 |
| "waste heat" AND (urban OR city OR building) AND (atmosphere OR climate OR heat island OR urban heat island) | 303 | 15 |
| "metabolic heat" AND (urban OR city OR atmosphere) | 33 | 27 |
| ("heat emission" OR "heat emissions") AND (urban OR city OR atmosphere OR anthropogenic) | 283 | 85 |
| "human-induced heat" | 6 | 5 |
| "anthropogenic moisture" | 8 | 5 |
| *Net addition from the additional branches* | | +220 |
| *Finalized query output (before de-duplication and cleanup)* | **3,038** | |

**TABLE S2. Study-level sensitivity values used in the anthropogenic heat sensitivity figure. Study labels use only the first author for simplicity, with initials added where needed to distinguish duplicate labels. The column $n$ is the number of estimates extracted from each study.**

| Study label | Title | DOI | Type | $n$ | Geometric mean K (W m$^{-2}$)$^{-1}$ | Range |
|---|---|---|---|---|---|---|
| Ghadban (2020) | A novel method to improve temperature forecast in data-scarce urban environments with application to the urban heat island in Beirut | `https://doi.org/10.1016/j.uclim.2020.100648` | forcing-based | 1 | 0.002 | 0.002 |
| Cao (2017) | Numerical and experimental study of sensitivity factors on heat island of residence community | `https://doi.org/10.1007/s10652-017-9544-x` | forcing-based | 1 | 0.00233 | 0.00233 |
| Nakajima (2021) | Human behaviour change and its impact on urban climate: restrictions with the G20 Osaka Summit and COVID-19 outbreak | `https://doi.org/10.1016/j.uclim.2020.100728` | effective | 2 | 0.00245 | 0.002–0.003 |
| Lin (2008) | Urban heat island effect and its impact on boundary layer development and land-sea circulation over northern Taiwan | `https://doi.org/10.1016/j.atmosenv.2008.03.015` | forcing-based | 1 | 0.003 | 0.003 |
| Khanh (2025) | Impact of anthropogenic heat on air temperature: a first-order estimate using dimensional analysis and numerical simulations | `https://doi.org/10.1029/2024GL114400` | forcing-based | 2 | 0.005 | 0.001–0.025 |
| McCarthy (2010) | Climate change in cities due to global warming and urban effects | `https://doi.org/10.1029/2010GL042845` | forcing-based | 2 | 0.00632 | 0.003–0.01 |
| Mussetti (2022) | Do electric vehicles mitigate urban heat? The case of a tropical city | `https://doi.org/10.3389/fenvs.2022.810342` | forcing-based | 2 | 0.00663 | 0.004–0.011 |
| Huang (2021) | The synergistic effect of urban heat and moisture islands in a compact high-rise city | `https://doi.org/10.1016/j.buildenv.2021.108274` | effective | 1 | 0.00678 | 0.00678 |
| Chen (2016) | Model analysis of urbanization impacts on boundary layer meteorology under hot weather conditions: a case study of Nanjing, China | `https://doi.org/10.1007/s00704-015-1535-6` | forcing-based | 1 | 0.0069 | 0.0069 |
| Xie (2024) | Could residential air-source heat pumps exacerbate outdoor summer overheating and winter overcooling in UK 2050s climate scenarios? | `https://doi.org/10.1016/j.scs.2024.105811` | effective | 2 | 0.00748 | 0.004–0.011 |
| Bueno (2012) | A resistance-capacitance network model for the analysis of the interactions between the energy performance of buildings and the urban climate | `https://doi.org/10.1016/j.buildenv.2012.01.023` | effective | 1 | 0.0075 | 0.005–0.01 |
| Karlicky (2026) | Sensitivity of modeled urban climate to urban canopy parameters over central Europe | `https://doi.org/10.1088/2752-5295/ae228c` | forcing-based | 2 | 0.00792 | 0.0057–0.011 |
| Wang (2023) | Quantifying the impacts of high-resolution urban information on the urban thermal environment | `https://doi.org/10.1029/2022JD038048` | forcing-based | 1 | 0.00813 | 0.00813 |

*Continued on next page*

| **Study label** | **Title** | **DOI** | **Type** | $n$ | **Geometric mean** $\mathbf{K\ (W\ m^{-2})^{-1}}$ | **Range** |
|---|---|---|---|---|---|---|
| Xin (2023) | Study of urban thermal environment and local circulations of Guangdong-Hong Kong-Macao Greater Bay Area using WRF and local climate zones | `https://doi.org/10.1029/2022JD038210` | effective | 2 | 0.00986 | 0.0021–0.024 |
| Yamaguchi (2025) | Urban cooling and CO2 reduction potentials of mass deployment of heat pump water heaters in Tokyo | `https://doi.org/10.1016/j.uclim.2025.102374` | effective | 2 | 0.0109 | 0.007–0.017 |
| Sun (2026) | Modeling urban traffic heat flux in the Community Earth System Model: formulation and validation for two test sites | `https://doi.org/10.1029/2025MS005435` | effective | 4 | 0.0128 | 0.008–0.019 |
| Bueno (2011) | Combining a detailed building energy model with a physically based urban canopy model | `https://doi.org/10.1007/s10546-011-9620-6` | effective | 2 | 0.014 | 0.013–0.015 |
| Gutman (1975) | Response of the urban boundary layer to heat addition and surface roughness | `https://doi.org/10.1007/bf00215641` | forcing-based | 1 | 0.014 | 0.014 |
| Chen (2024) | Modelling the impact of building energy consumption on urban thermal environment: the bias of the inventory approach | `https://doi.org/10.1016/j.uclim.2023.101802` | effective | 1 | 0.015 | 0.015 |
| Li XC (2024) | Elevated urban energy risks due to climate-driven biophysical feedbacks | `https://doi.org/10.1038/s41558-024-02108-w` | effective | 1 | 0.0155 | 0.001–0.03 |
| González-Aparicio (2014) | Impact of city expansion and increased heat fluxes scenarios on the urban boundary layer of Bilbao using Enviro-HIRLAM | `https://doi.org/10.1016/j.uclim.2014.07.010` | forcing-based | 2 | 0.0168 | 0.0125–0.0225 |
| Sorbjan (1982) | Some numerical urban boundary-layer studies | `https://doi.org/10.1007/BF00124707` | forcing-based | 1 | 0.017 | 0.017 |
| Hu (2012) | Numerical investigation on the urban heat island in an entire city with an urban porous media model | `https://doi.org/10.1016/j.atmosenv.2011.09.064` | forcing-based | 1 | 0.017 | 0.017 |
| Fan (2005) | Modeling the impacts of anthropogenic heating on the urban climate of Philadelphia: a comparison of implementations in two PBL schemes | `https://doi.org/10.1016/j.atmosenv.2004.09.031` | forcing-based | 8 | 0.018 | 0.002–0.08 |
| Ko (2026) | Modeling the distribution, impacts, and mitigation of anthropogenic heat in Los Angeles | `https://doi.org/10.1029/2026JD046326` | forcing-based | 3 | 0.0182 | 0.01–0.03 |
| Yu (1975) | Numerical study of the nocturnal urban boundary layer | – | forcing-based | 1 | 0.0185 | 0.017–0.02 |
| Tao (2022) | Impact of anthropogenic heat emissions on meteorological parameters and air quality in Beijing using a high-resolution model simulation | `https://doi.org/10.1007/s11783-021-1478-3` | forcing-based | 4 | 0.0187 | 0.00625–0.0314 |
| Feng (2014) | Impact of anthropogenic heat release on regional climate in three vast urban agglomerations in China | `https://doi.org/10.1007/s00376-013-3041-z` | forcing-based | 3 | 0.0201 | 0.0169–0.0235 |
| Salvati (2019) | Climatic performance of urban textures: analysis tools for a Mediterranean urban context | `https://doi.org/10.1016/j.enbuild.2018.12.024` | forcing-based | 2 | 0.0202 | 0.011–0.037 |
| Myrup (1969) | A numerical model of the urban heat island | – | forcing-based | 1 | 0.023 | 0.023 |
| Lei (2022) | Numerical study of micro-thermal environment in block based on porous media model | `https://doi.org/10.3390/buildings12050595` | forcing-based | 1 | 0.025 | 0.025 |
| Neunhäuserer (2007) | Towards urbanisation of the non-hydrostatic numerical weather prediction model Lokalmodell (LM) | `https://doi.org/10.1007/s10546-007-9159-8` | forcing-based | 2 | 0.0283 | 0.016–0.05 |

*Continued on next page*

| Study label | Title | DOI | Type | $n$ | Geometric mean | Range |
|---|---|---|---|---|---|---|
| | | | | | K (W m$^{-2}$)$^{-1}$ | |
| Li (2013) | A multi-resolution ensemble study of a tropical urban environment and its interactions with the background regional atmosphere | https://doi.org/10.1002/jgrd.50795 | forcing-based | 2 | 0.0287 | 0.006–0.12 |
| Aoyagi (2012) | Numerical simulation of the surface air temperature change caused by increases of urban area, anthropogenic heat, and building aspect ratio in the Kanto-Koshin Area | https://doi.org/10.2151/jmsj.2012-B02 | forcing-based | 2 | 0.0292 | 0.0255–0.0334 |
| Swaid (1991) | Thermal effects of artificial heat sources and shaded ground areas in the urban canopy layer | https://doi.org/10.1016/0378-7788(90)90137-8 | forcing-based | 1 | 0.033 | 0.033 |
| Block (2004) | Impacts of anthropogenic heat on regional climate patterns | https://doi.org/10.1029/2004GL019852 | forcing-based | 2 | 0.0335 | 0.015–0.075 |
| Atwater (1972) | Thermal effects of urbanization and industrialization in the boundary layer: a numerical study | https://doi.org/10.1007/bf02033921 | forcing-based | 4 | 0.0395 | 0.012–0.11 |
| Li D (2024) | Structural uncertainty in the sensitivity of urban temperatures to anthropogenic heat flux | https://doi.org/10.1029/2024MS004431 | forcing-based | 1 | 0.0415 | 0.003–0.08 |
| Wang (2019) | A modified building energy model coupled with urban parameterization for estimating anthropogenic heat in urban areas | https://doi.org/10.1016/j.enbuild.2019.109377 | effective | 1 | 0.045 | 0.04–0.05 |
| Lyu (2024) | Factors influencing the spatial variability of air temperature urban heat island intensity in Chinese cities | https://doi.org/10.1007/s00376-023-3012-y | effective | 1 | 0.058 | 0.058 |
| Swaid (1993) | Urban climate effects of artificial heat sources and ground shadowing by buildings | https://doi.org/10.1002/joc.3370130707 | forcing-based | 1 | 0.064 | 0.064 |
| Kikegawa (2022) | A quantification of classic but unquantified positive feedback effects in the urban-building-energy-climate system | https://doi.org/10.1016/j.apenergy.2021.118227 | effective | 3 | 0.0804 | 0.016–0.369 |
| Ming (2021) | Numerical investigation on the urban heat island effect by using a porous media model | https://doi.org/10.3390/en14154681 | forcing-based | 1 | 0.083 | 0.083 |
| Jacobson (2014) | Effects of biomass burning on climate, accounting for heat and moisture fluxes, black and brown carbon, and cloud absorption effects | https://doi.org/10.1002/2014JD021861 | forcing-based | 1 | 1.26 | 1.24–1.28 |



\end{document}